\RequirePackage{fix-cm}
\documentclass[Afour,sageh,times]{sagej}
\usepackage{amsmath}
\usepackage{amssymb}
\usepackage{algorithm}
\usepackage{pgfplots}
\usepackage{graphicx}
\usepackage{calc}
\usepackage{mathtools}
\usepackage{color}
\usepackage[inline]{enumitem}
\usepackage{float}
\usepackage{subfigure}
\usepackage{tikz}
\usepackage{xfrac}
\usepackage[hidelinks]{hyperref}
\usepackage{algpseudocode}
\usepackage{longtable}
\usepackage{booktabs}
\usepackage{xspace}
\usepackage{manfnt}
\usepackage{siunitx}
\usepackage{fmtcount}
\usepackage[normalem]{ulem}
\usepackage{commands}

\begin{document}

\title{%
A fine-grained parallelization of the immersed boundary method
}

\author{%
Andrew Kassen\affilnum{1},
Varun Shankar\affilnum{2} and
Aaron L. Fogelson\affilnum{1,3}
}

\affiliation{%
\affilnum{1}Department of Mathematics, University of Utah, Salt Lake City, UT, USA\\
\affilnum{2}School of Computing, University of Utah, Salt Lake City, UT, USA\\
\affilnum{3}Department of Biomedical Engineering, University of Utah, Salt Lake City, UT, USA
}

\corrauth{Aaron L. Fogelson, 155 S 1400 E, JWB 233, Salt Lake City, UT 84112, USA}
\email{fogelson@math.utah.edu}

\begin{abstract}
We present new algorithms for the parallelization of Eulerian-Lagrangian interaction
operations in the immersed boundary method. Our algorithms rely on two well-studied
parallel primitives: key-value sort and segmented reduce. The use of these parallel
primitives allows us to implement our algorithms on both graphics processing units (GPUs)
and on other shared memory architectures. We present strong and weak scaling tests on
problems involving scattered points and elastic structures. Our tests show that our
algorithms exhibit near-ideal scaling on both multicore CPUs and GPUs\@.
\end{abstract}


\maketitle

\section{Introduction}

Many problems in biophysics involve the interaction of an incompressible fluid and an
immersed elastic interface. The solution to these problems can be approximated using the
immersed boundary (IB) method, which was developed in~\cite{Peskin:1972wa} to simulate
blood flow through the heart. The relative ease by which the IB method can be
incorporated into a Navier-Stokes solver has led to its popularity in myriad applications
(see, e.g.~\cite{Iaccarino:2005ii,Griffith:2020hi} and references therein). The IB method
couples the equations governing fluid velocity and pressure to those governing the
interface movement and elastic forces via operations we will refer to as
\term{interpolation} and \term{spreading}. Fluid velocities are interpolated to points on
the interface, and forces on the interface are spread to the fluid. Fluid properties are
discretized on a fixed Eulerian grid. The interface is represented by a set of mobile
Lagrangian points. This presents a problem for effective parallelization of interpolation
and spreading, as these operators must be reconstructed at each timestep to account for
the motion of the Lagrangian points relative to the Eulerian grid. The Eulerian and
Lagrangian points, on the other hand, can be treated as fixed in their respective
coordinate spaces for several timesteps, if not for the entirety of a simulation. We may
therefore treat solving the fluid equations and computing elastic forces as {\reva the result of
a black-box computation} and focus primarily on the interpolation and spreading
operations.

We are interested in the case of neutrally buoyant, elastic immersed structures that move
at the local fluid velocity. In particular, we aim to simulate whole blood, which is
composed of red blood cells (RBCs) and platelets immersed in blood plasma, within a
vessel lined by endothelial cells. The cells are elastic, and as they deform with the
flow of the enveloping fluid, {\reva they} impose a force on the fluid. Approximately 40\% of
the volume in healthy human blood is occupied by RBCs, {\reva and domains containing just a few
cells} may require tens or hundreds of thousands of points to discretize these cells for
use within the IB method.

To take advantage of modern computing architectures, with ever-increasing numbers of
processors, it is necessary to develop parallel algorithms for the IB method.~%
\cite{McQueen:1997kw} present a domain decomposition scheme to parallelize the
interpolation and spreading operations on the Cray C-90 computer with shared memory and a
modest number of vector processors. Their results illustrate the need for a fast
interpolation and spreading: even parallelized, they spend roughly half of the wall clock
time spreading and interpolating.~\cite{Fai:2013do} adapted this domain decomposition
scheme for use on a general purpose graphical processing unit (GPGPU; GPU for short).~%
\cite{Patel:2012tc} parallelized spreading on a GPU by processing one Lagrangian point at
a time. Because neither of these approaches can concurrently process arbitrary Lagrangian
points, it is easy to find cases for which they perform poorly. An alternative is to
distribute work among several devices, each with its own memory. This idea underpins the
popular IBAMR library (see~\cite{Griffith:2007uk,Griffith:2007do,Griffith:2009gg,
Griffith:2011gi,Griffith:2017id}), which also adaptively refines the mesh around the
immersed structure. With proper load-balancing, this allows a serial algorithm to process
a smaller portion of work. A cluster of multicore devices, however, will not be used
effectively without a shared-memory parallel algorithm. The cuIBM (\cite{Layton:2011um})
and PetIBM (\cite{Mesnard:2017te,Chuang:2018ej}) libraries implement an adaptive IB
method for prescribed motion on single- and multi-GPU architectures, respectively. The
authors demonstrate their method on a few two-dimensional test problems. Their
implementation explicitly constructs the spreading and interpolation operators, which are
sparse. {\reva The interpolation matrix has approximately the same number of nonzero
entries per row, can be constructed in parallel in compressed sparse row (CSR) format,
and its multiplication against a column vector parallelizes easily as a series of sparse
dot products. The spreading matrix, with approximately the same number of nonzero
entries per column, can be constructed in parallel in compressed sparse column (CSC)
format, for which parallel multiplication against a column vector typically requires
conversion to a different format. Even after conversion to CSR, the spreading matrix does
not have approximately the same number of entries per row so does not always yield
effective parallelization of matrix-vector products.}

GPUs have restrictions on their parallelization. They use single instruction,
multiple data (SIMD) parallelism, in which each computational unit, or thread, executes
the same instruction on its own data. Concurrency on the GPU is typically limited by the
amount of shared memory, which is shared among a group of threads, and register memory,
which is accessible only to a single thread. The alternative is to use global memory,
which is slow in general, but faster when accesses are sequential (or ``coalesced'').
These restrictions on the GPU imply that an effective algorithm for the GPU translates
well to other shared-memory architectures, such such as multicore CPUs, which support
parallelism, SIMD instructions, advanced instruction pipelining, and out-of-order
execution. We therefore develop parallel spreading and interpolation algorithms
applicable to both GPUs and multicore CPUs. The success of our new algorithms relies in
dividing these operations into trivially parallelizable tasks and parallel primitives.

The remainder of this paper is organized into four sections. Section~\ref{sec:ib} gives
an overview of the IB method, and describes the role of the interpolation and spread
operations. In Section~\ref{sec:parallel}, we discuss the parallelization of
interpolation and introduce a new parallelization for the spreading operation. In
Section~\ref{sec:results}, we demonstrate that the concurrency of these algorithms scales
with the number of IB points, independently of the Eulerian grid. We confirm the
convergence of the IB method with these new algorithms. We illustrate the suitability of
our algorithms to both GPUs and multicore CPU architectures with both weak and strong
scaling tests. Finally, we summarize our findings in Section~\ref{sec:summary}.


\section{Review of the immersed boundary method} \label{sec:ib}

Consider a $d$-dimensional ($d=2$ or 3) rectangular domain $\domain$, which is filled
with a viscous incompressible fluid with constant viscosity $\mu$ and density $\rho$, and
contains an immersed elastic structure, $\interface$. The structure is impermeable to the
fluid and moves at the local fluid velocity, is deformed by this motion, and imparts a
force to the fluid. Otherwise, the interface is treated as part of the fluid.

The fluid velocity, $\u = \u(\x,\,t)$, and pressure, $p = p(\x,\,t)$, are governed by the
incompressible Navier-Stokes equations for a Newtonian fluid,
\begin{gather}
    \label{eq:ins-evol}
    \rho(\u_t + \u\cdot\grad\u) = \mu\Delta\u - \grad p + \f, \\
    \label{eq:ins-incomp}
    \div\u = 0,
\end{gather}
where $\f$ is the elastic force density. This is a set of $d+1$ equations in $d+1$
unknowns: the $d$ components of $\u$, and $p$. The equations are written relative to the
Eulerian frame, so that the coordinates $\x$ are independent variables. Throughout, we
write Eulerian quantities in the lowercase Latin alphabet.

Let $\X=\X(\params,\,t)$ represent a parametrization of the Cartesian coordinates of the
immersed interface with material coordinates $\params$ at time $t$. Let $\L[\X]$ be the
energy density functional for the elastic interface material. The elastic force density
is computed by evaluating
\begin{equation}
    \label{eq:forces}
    \F = -\delta \L[\X],
\end{equation}
where $\delta$ represents the first variation. Uppercase Latin letters represent
Lagrangian quantities and are functions of $\params$ and $t$.

The interface moves at the local fluid velocity, and force balance on the interface
between the interface and fluid dictates that the interface force on the fluid is equal
to the elastic force. Analytically, the fluid-interface interactions can be written
\begin{gather}
    \label{eq:interpolation}
    \U(\params,\,t) = \int_\domain \Dirac(\x-\X(\params,\,t)) \u(\x,\,t)\d\x,\ \text{and} \\
    \label{eq:spreading}
    \f(\x,\,t) = \int_\interface \Dirac(\x-\X(\params,\,t))\F(\params,\,t)\d\params,
\end{gather}
where $\U(\params,\,t)$ represents the derivative of $\X(\params,\,t)$ with respect to
$t$, and $\Dirac(\x-\X(\params,\,t))$ is the Dirac $\Dirac$-function centered at
$\X(\params,\,t)$. Equation~\eqref{eq:interpolation} is called \term{interpolation}, and
the result of the right-hand side is the fluid velocity at $\X(\params,\,t)$; namely,
$\U(\params,\,t) = \u(\X(\params,\,t),\,t)$. Equation~\eqref{eq:spreading} is called
\term{spreading}. {\reva$\F(\params,\,t)$ and $\f(\x,\,t)$ may not have the same units, but
the force $\F(\params,\,t)\d\params$ over $\d\params$ is ``spread'' to the force
$\f(\x,\,t)\d\x$ over $\d\x$.}

The fluid equations~\eqref{eq:ins-evol}--\eqref{eq:ins-incomp} are spatially discretized
on a regular background grid of spacing $h$ so that $\domain$ is divided into square or
cubic cells of side length $h$. Because of the checkerboard instability (see, e.g.,~%
\cite{Wesseling:2001ci}) in solving the Navier-Stokes equations on collocated regular
grids, {\revab we use a MAC grid (\cite{Welch:1965jv}) to stagger the components of Eulerian
vector quantities. For example, if $\vec{v}$ is an Eulerian vector-valued function,
$\e$ is a canonical basis vector, and $\x$ is a cell center, $\e\cdot\vec{v}$ is
discretized at $\x - h\e/2$. These discretization points comprise a new, regular
grid with spacing $h$. For the component under consideration, $\domain_h$ denotes the set
of discretization points with cardinality $n_\omega$.} The momentum equation~%
\eqref{eq:ins-evol} is discretized in time using a {\reva first-order method based on
the} implicit-explicit Runge-Kutta method described in~\cite{Peskin:2002go}, and the
incompressibility condition~\eqref{eq:ins-incomp} is satisfied using PmIII of~%
\cite{Brown:2001bq}.

The Lagrangian force density~\eqref{eq:forces} is evaluated at a set of points, usually
{\revb with \emph{fixed} material coordinates}, referred to as IB points. The notation
$\X_j=\X(\params_j,\,t)$ refers to an individual IB point on $\interface$ in Cartesian
coordinates. The typical heuristic for distributing the points $\X_j$ on a connected
interface places neighboring IB points at most $h$ apart from one another, and often at
most $h/2$ apart. We therefore denote the set of IB points by $\interface_h$ and the
number of IB points by $n_\gamma$. From these points, we construct a smooth approximation
to the interface using radial basis functions (RBFs), as in~\cite{Shankar:2015km}. This
allows us to calculate geometric properties using well-known formulas, and evaluate the
forces~\eqref{eq:forces} in analytic form, similar to that of~\cite{Maxian:2018ek}.

The singular integrals~\eqref{eq:interpolation} and~\eqref{eq:spreading} do not lend
themselves easily to evaluation. In particular, it is unlikely that IB points and grid
points will coincide. For a regular grid with spacing $h$, we replace the Dirac
$\Dirac$-function with a regularized kernel, $\Dirac_h$, which is a product of
one-dimensional kernels, $h^{-1}\kernel(h^{-1}x)$.~\cite{Griffith:2020hi} gives several
options for $\kernel$. {\revb The methods described in this paper do not depend on a particular
$\kernel$, but we restrict ourselves to the cosine kernel from~\cite{Peskin:2002go}.
The cosine kernel is simple to implement, but does not satisfy all of the moment
conditions for a 4-point kernel, though it approximates a kernel with the appropriate
discrete properties.}

Let $\domain_h$ be one of the Eulerian grids for vector-valued quantities. {\revb Discretizing
the component of equations~\eqref{eq:interpolation} and~\eqref{eq:spreading} on
$\domain_h$ and $\interface_h$, respectively, corresponding to canonical basis vector
$\e$ yields
\begin{align}
    \label{eq:disc-interp}%
    \e\cdot\U^n_j &= \sum_{i=1}^{n_\omega} \delta_h(\x_i-\X^n_j)\e\cdot\u^n_i h^d \quad \text{and} \\
    \label{eq:disc-spread}%
    \e\cdot\f^n_i &= \sum_{j=1}^{n_\gamma} \delta_h(\x_i-\X^n_j)\e\cdot\F^n_j \d\theta_j,
\end{align}
where $\u^n_i$ and $\f^n_i$ are discrete approximations to $\u$ and $\f$}
at $\x_i\in\domain_h$ and time $t=t_n$, and {\revb$\U^n_{\smash j}$} and $\F^n_{\smash j}$
are their Lagrangian counterparts at $\X_{\smash j}$, respectively. The terms $h^d$ and
$\d\theta_{\smash j}$ are integration weights analogous to $\d\x$ and $\d\params$. For
a topologically spherical interface, we compute weights on the unit sphere using RBFs,
using a simplified version of the method described in~\cite{Fuselier:2013coba}, and use
the Jacobian of the mapping between the sphere and interface to obtain
$\d\theta_{\smash j}$. Both~\eqref{eq:disc-interp} and~\eqref{eq:disc-spread} look like a
matrix-vector multiplication, so we define the \emph{spreading matrix}
$\spread=(\delta_h(\x_i-\X^n_{\smash j}))$ with row $i$ and column $j$. We call its
transpose, $\interp$, the \emph{interpolation matrix}. Collecting values
{\reva $\e\cdot\u^n_ih^d$ at each grid point as $\vec{v}^n$ and
$\e\cdot\F^n_{\smash j}\d\theta_{\smash j}$ at each IB point as $\vec{G}^n$}, we
rewrite equations~\eqref{eq:disc-interp} and~\eqref{eq:disc-spread} in matrix form as
{\reva
\begin{align}
    \label{eq:matrix-interp}%
    \U^n &= \interp{\vec{v}}^n \quad\text{and} \\
    \label{eq:matrix-spread}%
    \f^n &= \spread{\vec{G}}^n,
\end{align}}
respectively.

Collectively, the equations~\eqref{eq:ins-evol}--\eqref{eq:spreading} constitute the
IB method, and while different implementations {\reva of the IB method use different
spatial and temporal discretizations, surface representations, elastic models, or choice
of $\delta_h$}, a single step proceeds with timestep $k$ roughly as follows:
\begin{enumerate}[label=(\texttt{\alph*}), itemsep=0ex]
    \item interpolate $\u^n$ to $\X^n$ to get {\revb $\U^\ast$},
    \item predict Lagrangian positions $\X^\ast = \X^n + k{\revb \U^\ast}$,
    \item compute Lagrangian forces $\F^\ast$ using positions $\X^\ast$,
    \item spread {\reva$\F^\ast$} from $\X^\ast$ to get $\f^\ast$,
    \item solve for updated Eulerian velocities $\u^{n+1}$,
    \item interpolate $\u^{n+1}$ to $\X^n$ to get {\revb$\U^{n+1}$}, and
    \item update $\X^{n+1} = \X^n + k {\revb\U^{n+1}}$.
\end{enumerate}
We group these steps into 3 categories: the purely Eulerian (\texttt{e}); the purely
Lagrangian (\texttt{b}), (\texttt{c}), and (\texttt{g}); and the Eulerian-Lagrangian
coupling (\texttt{a}), (\texttt{d}), and (\texttt{f}). We discuss the first two
categories in a forthcoming paper. The remainder of this paper discusses the
implementation and parallelization of the third.


\section{Parallelization strategies for interpolation and spreading} \label{sec:parallel}

Here, we describe a method for evaluating the discrete counterparts to
\begin{align}
    \label{eq:scalar-interp}
    E(\X) &= \int_\domain \Dirac_h(\x-\X)e(\x) \d\omega \quad \text{and}\\
    \label{eq:scalar-spread}
    \ell(\x) &= \int_\interface \Dirac_h(\x-\X)L(\X) \d\gamma
\end{align}
where scalar-valued Eulerian function $e:\domain\to\mathbb{R}$ is interpolated to
Lagrangian point $\X$ and Lagrangian function $L:\interface\to\mathbb{R}$ is spread to
Eulerian point $\x$. For vector-valued functions, such as the Eulerian fluid velocity
$\u$ and Lagrangian force density $\F$, the algorithm can be applied to each component
individually. For a staggered grid, this is necessary, as the grid for each component has
a different set of grid points, and may have a different number of grid points. In the
following, we discuss parallelizing equation \eqref{eq:scalar-interp} and introduce a
novel parallelization scheme for evaluating equation \eqref{eq:scalar-spread}. We begin
by defining some notation that will be used throughout the description of these
algorithms.

\subsection{Preliminaries}

For a regular grid on $\domain$ with grid spacing $h$, $\domain_h$, any grid point
$\x\in\domain_h$ can be decomposed into a vector $\vec{i}$ with integer components and a
fixed grid staggering, $\stag\in[0,\,1)^d$, such that $\x=h(\vec{i}+\stag)$. Let
$\interface_h$ be a discretization of the interface $\interface$. For the purposes of
interpolation and spreading, we can think of $\interface_h$ as a collection of arbitrary
points in $\domain$.

\begin{figure}[t]
    \begin{center}
    \begin{tikzpicture}[scale=1.25]
        \draw[help lines] (-0.1, -0.1) grid (5.1, 5.1);
        \draw[color=gray, fill=gray, opacity=0.2] (0.6, 0.6) rectangle (4.6, 4.6);
        \draw[pattern={thick horizontal lines}, pattern color=gray, draw=none] (2, 2.5) rectangle (3, 3.5);
        \draw[thick, gray, fill=none] (2, 2.5) rectangle (3, 3.5);
        \draw[pattern=thick vertical lines, pattern color=gray, draw=none] (2.5, 2) rectangle (3.5, 3);
        \draw[thick, gray, fill=none] (2.5, 2) rectangle (3.5, 3);
        \node[circle, color=black, fill=black, inner sep=2pt] at (2.6, 2.6) {};

        \foreach \x in {0,...,5}
            \foreach \y in {0,...,4}
                \node[regular polygon, regular polygon sides=3, color=black, fill=none, inner sep=1.414pt, shape border rotate=-90, thick, draw] at (\x, \y+0.5) {};
        \foreach \x in {0,...,4}
            \foreach \y in {0,...,5}
                \node[regular polygon, regular polygon sides=3, color=black, fill=none, inner sep=1.414pt, thick, draw] at (\x+0.5, \y) {};
        \node[regular polygon, regular polygon sides=3, color=black, fill=black, inner sep=1.414pt, shape border rotate=-90] at (2, 2.5) {};
        \node[regular polygon, regular polygon sides=3, color=black, fill=black, inner sep=1.414pt] at (2.5, 2) {};
        \node[black] at (0.25, 4.75) {\sffamily(a)};
    \end{tikzpicture}
    
    \vspace{1em}

    \begin{tikzpicture}[scale=1.25]
        \draw[help lines] (-0.1, -0.1) grid (5.1, 5.1);
        \draw[color=gray, fill=gray, opacity=0.2] (1.1, 1.1) rectangle (4.1, 4.1);
        \draw[pattern=thick horizontal lines, pattern color=gray, draw=none] (2.5, 2) rectangle (3.5, 3);
        \draw[thick, gray, fill=none] (2.5, 2) rectangle (3.5, 3);
        \draw[pattern=thick vertical lines, pattern color=gray, draw=none] (2, 2.5) rectangle (3, 3.5);
        \draw[thick, gray, fill=none] (2, 2.5) rectangle (3, 3.5);
        \node[circle, color=black, fill=black, inner sep=2pt] at (2.6, 2.6) {};

        \foreach \x in {0,...,5}
            \foreach \y in {0,...,4}
                \node[regular polygon, regular polygon sides=3, color=black, fill=none, inner sep=1.414pt, shape border rotate=-90, thick, draw] at (\x, \y+0.5) {};
        \foreach \x in {0,...,4}
            \foreach \y in {0,...,5}
                \node[regular polygon, regular polygon sides=3, color=black, fill=none, inner sep=1.414pt, thick, draw] at (\x+0.5, \y) {};
        \node[regular polygon, regular polygon sides=3, color=black, fill=black, inner sep=1.414pt, shape border rotate=-90] at (3, 2.5) {};
        \node[regular polygon, regular polygon sides=3, color=black, fill=black, inner sep=1.414pt] at (2.5, 3) {};
        \node[black] at (0.25, 4.75) {\sffamily(b)};
    \end{tikzpicture}
    \end{center}

    \caption{%
A region of a 2-dimensional domain containing point $\X$, indicated by a black
circle. Light gray lines indicate physical units in increments of $h$, and $\X$
has the same coordinates in each subfigure. Light gray boxes indicate the
support of $\Dirac_h(\cdot-\X)$. Subfigure (a) has $s=4$ support points in each
dimension, and (b) has $s=3$. The grid for horizontal vector components has
staggering $\stag=(0,\,0.5)$ and is marked by right-pointing triangles, whereas
upright triangles mark the grid for vertical vector components and has
staggering $\stag=(0.5,\,0)$. Those grid points within the gray boxes are also
support points of $\X$. The horizontally- and vertically-striped gray boxes
denote the grid cell containing $\X$ on the horizontal or vertical component
grid, respectively. The filled triangles mark the point $\x=h(\idx{\X}+\stag)$,
corresponding to shift $\shift=(0,\,0)$ on the appropriate grid. The positions
of the grid cells in (a) are typical of all even $s$, and the positions of the
grid cells in (b) are typical of all odd $s$.
    }
    \label{fig:grid}
\end{figure}
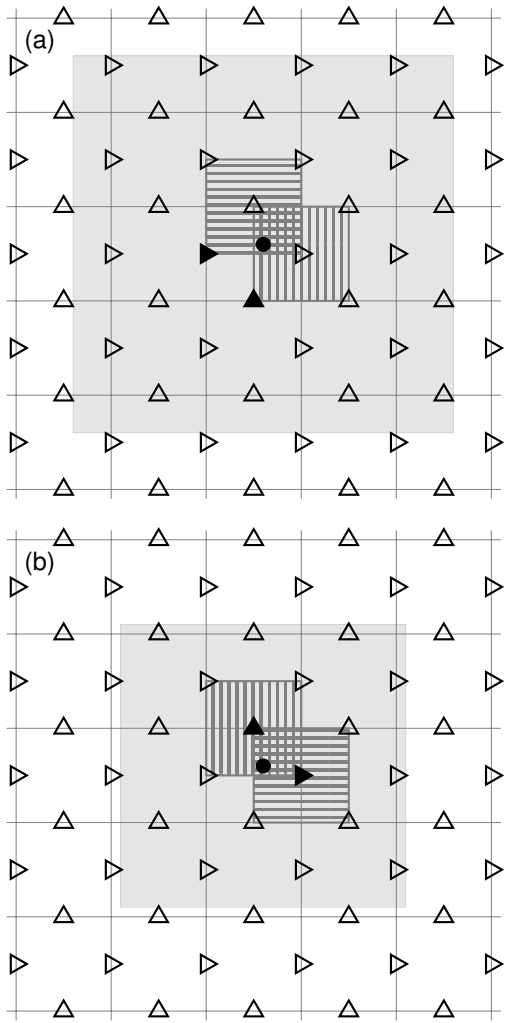

The kernel $\Dirac_h$ is the tensor product of scaled, one-dimensional kernels,
$h^{-1}\kernel(h^{-1}x)$. Let $\supp\kernel$ denote the support of $\kernel$ and define
\begin{equation}
    s[\kernel] = \max_{r\in[0,\,1)}|\supp\kernel(\cdot-r)\cap\mathbb{Z}|-1
\end{equation}
to be the size of the support in terms of unit intervals. For brevity, we write
$s=s[\kernel]$. For any $\X\in\domain$, there are at most $s^d$ grid points
$\x\in\domain_h$ for which $\Dirac_h(\x-\X)$ is nonzero. We call these points the
\term{support points} of $\X$. If $\X$ is sufficiently far from any boundary, {\revb the set of
points in $\domain$ with the same set of support points as $\X$ is a region of $\domain$
the same size as a background grid cell. These regions partition $\domain$ and therefore
contain at most one grid point.} For $\X$ near a non-periodic boundary, the region may be
smaller, and may not contain a grid point. We extend the grid so that each of these
regions contains a single grid point. {\revb For the MAC grid, this means adding grid
points to the boundary of $\domain$, and is only strictly necessary for cases where grid
points are not at a corner of a region, \latin{i.e.}, $s$ is odd}. We call this extension
$\bar{\domain}_h$. We consider these regions to be the \emph{de facto} grid cells, and
henceforth refer to them as such. The striped gray boxes in Figure~\ref{fig:grid}
therefore represent these cells, and illustrates the effect different choices of
$\kernel$ has on their location. We can now identify any $\X\in\domain$ with a grid point
$\x\in\bar{\domain}_h$ if they are in the same grid cell, and since $\x=h(\vec{i}+\stag)$
is the only grid point in the grid cell, we identify the cell by the integers $\vec{i}$.
We adopt the notation $\idx{\X}=\vec{i}$ for the function that maps points in $\domain$
to the vector of integers identifying the grid cell containing $\X$.

\begin{algorithm}
\caption{\revb Shift construction}
\label{algo:shifts}
\begin{algorithmic}[1]
\revb
\Procedure{shift}{$j,\,s,\,d$}
\State $\triangleright\ $\textbf{require}: $1 \le j \le s^d$
\State $\triangleright\ $\textbf{generate}: Shift $\shift_j$
\State $\shift\gets\vec{0}$ \Comment{$d$ zeros}
\For {$i=1,\,\ldots,\,d$}
    \State $\sigma_{i} \gets \operatorname{mod}(j-1,\,s) - \floor{s/2}$
    \State $j \gets \floor{(j-1)/s}$
\EndFor
\State \Return $\shift$
\EndProcedure
\end{algorithmic}
\end{algorithm}

We now turn our attention to the evaluation of $\Dirac_h(\x-\X)$. We assume
$\x\in\domain_h$ and write
\begin{equation}
    \label{eq:delta-defs}
    \begin{aligned}
        \Dirac_h(\x-\X)
        &\equiv \Dirac_h(h\shift - \Delta\x) \\
        &= \prod_{i=1}^d h^{-1}\kernel((\shift - h^{-1}\Delta\x)_i),
    \end{aligned}
\end{equation}
where $\Delta\x = \X-h(\idx{\X}+\stag)$ is the displacement of $\X$ from its associated
grid point, $\shift=\idx{\x}-\idx{\X}$, and subscript $i$ denotes the $i^\text{th}$
component. We refer to $\shift$ as a \term{shift}. Shifts that result in a possibly
nonzero value of $\Dirac_h$ are known \emph{a priori} based on the kernel $\kernel$, and
usually range from $-\floor{s/2}$ to $\floor{(s-1)/2}$ in each component. We can
therefore assign an order to the shifts, $\shift_1$, $\shift_2$, \dots, $\shift_{s^d}$.
{\revb For example, Algorithm~\ref{algo:shifts} computes $\shift_j$ given $s$ and $d$ in
colexicographic order.} We need only compute $\Delta\x$ once to be able to compute all
nonzero values of $\Dirac_h(\x-\X)$.

We need one more ingredient to construct $\spread$. We let $\x_k$ be the be the $k\th$
grid point, and for Eulerian function $e(\x)$, we construct the vector $\vec{e}$ with
$k\th$ entry $e_k = e(\x_k)$. Define the \term{grid indexing function}
\begin{equation}
    \label{eq:grid-index-fn}
    \#(\vec{i}) = \begin{cases}
        k, & \idx{\x_k}=\vec{i},\ \x_k\in\domain_h \\
        \error, & \text{otherwise}.
    \end{cases}
\end{equation}
The value $\error$ indicates an Eulerian point outside of $\domain$, does not have a
corresponding row in $\spread$, and therefore does not contribute to spreading or
interpolation. We are now ready to construct $\spread$. Let $\X_j$ be an IB
point. The $j\th$ column of $\mathcal{S}$ is zero except for up to $s^d$ values where,
for $i=1,\,\ldots,\,s^d$, if $\#(\idx{\X_j}+\shift_i)=k\neq\error$,
\begin{equation}
    \label{eq:s-columnwise}
    \spread_{kj} = \Dirac_h\left(h\left(\shift_i+\idx{\X_j}+\stag\right)-\X_j\right).
\end{equation}
From a practicality standpoint, it is unnecessary to explicitly construct $\spread$. We
illustrate this with the serial spreading algorithm.


\subsection{Serial spread}

\begin{algorithm}
\caption{Serial spread}
\label{algo:serial-spread}
\begin{algorithmic}[1]
\Procedure{seq-spread}{$\interface_h,\,\domain_h,\,\vec{L}$}
\State $\triangleright\ $\textbf{generate}: {\revb\parbox[t]{\dimexpr\linewidth-1.5em-\genlinewidth}{Approximate values of $\ell$ (Eq.~\eqref{eq:scalar-spread}) at each point in $\domain_h$}}
\For {$i = 1,\,\ldots,\,n_\gamma$}\Comment{Loop over IB points}
    \State $\x \gets h(\idx{\X_i}+\stag)$\Comment{$\X_i\in\interface_h$, $\x\in\domain_h$}\label{line:serial-spread-x}
    \State $\Delta\x \gets \x-\X_i$
    \For {$j = 1,\,\ldots,\,s^d$}\Comment{Loop over shifts}
        \State {\revb$\shift \gets \Call{shift}{j,\,s,\,d}$ \Comment{Algorithm~\ref{algo:shifts}}}
        \State $w \gets \Dirac_h(\Delta\x+h\shift)$
        \State $k \gets \#(\idx{\x} + \shift)$ \label{line:serial-spread-k}
        \If {$k\neq\error$}
            \State $\ell_k \gets \ell_k + w \cdot L_i$ \label{line:serial-spread-write}
        \EndIf
    \EndFor
\EndFor
\State \Return $\vec{\ell}$
\EndProcedure
\end{algorithmic}
\end{algorithm}

Algorithm \ref{algo:serial-spread} lists an example serial implementation of spreading in
pseudocode. The overall structure consists of two loops: a loop over the (indices of)
IB points, and a loop over the (indices of) shifts. From this, we see that for a fixed
choice of $\kernel$, and therefore $s$, the amount of work is $\bigo{n_\gamma}$, i.e.,
independent of the Eulerian grid. The spread values are accumulated into a vector
$\vec{\ell}$ (line \ref{line:serial-spread-write}). The target index, $k$ (line
\ref{line:serial-spread-k}), is computed using the grid indexing function, introduced in
the previous section. Note that $k$ depends on $\shift_j$, and $\x$, which in turn
depends on $\X_i$, as seen on line \ref{line:serial-spread-x}. This means that $k$
depends on both loop indices. There is no guarantee that unique pairs of the loop
variables $i$ and $j$ will yield unique target indices. As a result, simply parallelizing
one or both of the loops may lead to write contentions.

The property that runtime be independent of the Eulerian grid is desirable, as the
number of IB points is often much fewer than the number of grid points. In other words,
many grid cells will be empty of IB points, and unless they have nearby IB points, there
is no useful work to be done for that grid cell. An algorithm that depends on the
Eulerian grid will invariably involve wasted computational resources. We therefore aim to
preserve the independence property. This is achieved straightforwardly for interpolation.


\subsection{Parallelization of interpolation}

By construction, $\spread$ has approximately equal number of nonzero entries per column.
This means, that the interpolation matrix, $\interp$, has approximately equal number of
nonzero entries per row. This property is beneficial for parallelization. Consider the
$i\th$ row of $\interp$, which corresponds to interpolating to IB point $\X_i$ using the
values at its support points. There are at most $s^d$ values in that row, which
correspond to the shifts that give a potentially nonzero value for $\Dirac_h$. Compute
$\x=h(\idx{\X_i}+\stag)$. Then $\Delta\x=\x-\X_i$. Since the shifts $\{\shift_j\}$ are
known beforehand, we can compute $\Dirac_h(h\shift_j+\Delta\x)$ and accumulate products
\begin{equation*}
    E_i = \sum_{\substack{j=1\\\#(\shift_j+\idx{\x})\neq\error}}^{s^d}\Dirac_h(h\shift_j+\Delta\x)e_{\#(\shift_j+\idx{\x})},
\end{equation*}
where $\#(\vec{i})$ is defined in Equation \ref{eq:grid-index-fn}. This is done for each
IB point, for a total work proportional to the number of IB points.

\begin{algorithm}
\caption{Parallel interpolation}
\label{algo:par-interp}
\begin{algorithmic}[1]
\Procedure{par-interpolate}{$\interface_h,\,\domain_h,\,\vec{e}$}
\State $\triangleright\ $\textbf{generate}: {\revb\parbox[t]{\dimexpr\linewidth-1.5em-\genlinewidth}{Approximate values of $E$ (Eq.~\eqref{eq:scalar-interp}) at each point in $\interface_h$}}
\For {$i = 1,\,\ldots,\,n_\gamma$ \textbf{parallel}}
    \State $\x \gets h(\idx{\X_i}+\stag)$\Comment{$\X_i\in\interface_h$, $\x\in\domain_h$}
    \State $\Delta\x \gets \x-\X_i$
    \State $v \gets 0$\Comment{Accumulator}
    \For {$j = 1,\,\ldots,\,s^d$}
        \State {\revb $\shift \gets \Call{shift}{j,\,s,\,d}$ \Comment{Algorithm~\ref{algo:shifts}}}
        \State $w \gets \Dirac_h(\Delta\x+h\shift)$
        \State $k \gets \#(\idx{\x}+\shift)$
        \If {$k\neq\error$}
            \State $v \gets v + w \cdot e_k$\label{line:par-interp-acc}
        \EndIf
    \EndFor
    \State $E_i \gets v$\label{line:par-interp-write}
\EndFor
\State \Return $\vec{E}$
\EndProcedure
\end{algorithmic}
\end{algorithm}

Assigning one thread per IB point (i.e., one thread per row), this calculation can be
performed in parallel, and since the $i^\text{th}$ thread writes to the $i^\text{th}$
entry of $\vec{E}$, there are no write contentions. This can be seen on lines
\ref{line:par-interp-acc} and \ref{line:par-interp-write} in Algorithm
\ref{algo:par-interp}, where the accumulation happens in a temporary variable which is
ultimately written to $\vec{E}$. In this case, the target index depends only on the loop
variable $i$, and we can safely parallelize over this loop.  Because the number of
products is approximately the same for each row, each thread does approximately the same
amount of work. On architectures that enforce thread synchrony, such as GPUs, this means
that we do not incur a penalty from having threads wait for other threads to finish.

Since each thread computes the appropriate $\Dirac_h$-weights for its own row, it is
unnecessary to construct $\interp$ explicitly. Except allocating memory for $\vec{E}$,
all of the work for this algorithm is parallel, so we expect to see near-perfect scaling.
Additionally, other than using the grid spacing $h$ for scaling in various places, the
evaluation of $\Dirac_h$, and information about boundaries, there is no dependence on the
Eulerian grid. We now show how these properties can be maintained for the spreading
operation.


\subsection{Parallel spreading}

We return now to spreading. Consider a fixed $j$ in Algorithm \ref{algo:serial-spread}
so $\shift_j$ is fixed. If every IB point inhabits its own grid cell, the support point
for each IB point corresponding to $\shift_j$ is unique. In this case, we can spread to
those support points without concern for write contentions. This is unlikely to occur in
practice. On the other hand, if every IB point were in the same grid cell, values could
be accumulated in parallel before being spread. This too is unlikely to occur in
practice. We can instead employ the \emph{segmented reduce} algorithm, which, given a
list of values and a corresponding list of keys, will sum (reduce) consecutive values as
long as their keys match. The result is a potentially shorter list of reduced values and
a list of non-repeating keys, though they may not be unique within the list. Suppose we
were able to order the keys and values so that repeated keys are listed consecutively.
The result of reducing this sorted data is a list of unique keys and a list where all
values with the same key are combined. {\reva These observations lead us to assign a
unique index to each grid cell, and, for each IB point, to use the index for the grid
cell an IB point inhabits as its key. The list of keys is therefore identical for each
shift. If all IB points in the same grid cell are listed consecutively, \latin{i.e.}, all
occurrences of key $k$ in the list of keys are listed consecutively, then segmented
reduce accumulates values spread from all IB points within the same grid cell for the
given shift. This results in} at most one value per grid cell, and we can write without
contentions.

To ensure that repeated keys are listed consecutively, we use \emph{key-value sort},
which, given a list of values and a corresponding list of keys, sorts values according to
a partial ordering imposed on the keys. The result is a sorted list of keys and a
permuted list of values, but key-value pairs remain unchanged. {\reva That the keys are
ordered is not important, but all instances of a given key are grouped as a side-effect}. Computing keys and values,
sorting by key, and applying the segmented reduce algorithm once per shift spreads all
values. But notice that sorting once per shift results in the same sorted list of keys
each time. {\reva This also implies that values are permuted in the same manner for each
shift.} Instead of sorting values, we construct a permutation by sorting the indices of
the IB points. {\reva We now need only sort} once per spread operation, and using the
permutation, we can construct the list of values in sorted order for each shift. Now, we
apply segmented reduction to the sorted list of keys and newly constructed list of
values. {\reva Because the order of the keys dictates the code path taken by segemented
reduce, reduction proceeds identically for each shift, but with different data. We
discuss this further in the next section.} Finally, we write the reduced values to the
output. Computing values, reducing, and writing for each shift completes the calculation.

\begin{algorithm}
\caption{Parallel spread}
\label{algo:par-spread}
\begin{algorithmic}[1]
\Procedure{par-spread}{$\interface_h,\,\domain_h,\,\vec{L}$}
\State $\triangleright\ $\textbf{generate}: {\revb\parbox[t]{\dimexpr\linewidth-1.5em-\genlinewidth}{Approximate values of $\ell$ (Eq.~\eqref{eq:scalar-spread}) at each point in $\domain_h$}}
\For {$i = 1,\,\ldots,\,n_\gamma$ \textbf{parallel}} \label{line:par-spread-ps}
    \Statex \Comment{Loop over IB points}
    \State $K_i \gets \key(\idx{\X_i})$ \Comment{Sort key}
    \State $P_i \gets i$ \Comment{Initial ordering}
\EndFor
\State \textbf{sort} $\{P_i\}$ \textbf{by} $\{K_i\}$ \label{line:par-spread-sort} \Comment{$i\to P_i$ is a permutation} 
\State $q \gets \text{\textbf{count unique} }\{K_i\}$ \label{line:par-spread-q}
\For  {$j = 1,\,\ldots,\,s^d$} \label{line:par-spread-shifts} \Comment{Loop over shifts}
    \State $\{K'_i\} \gets \{K_i\}$
    \State {\revb $\shift \gets \Call{shift}{j,\,s,\,d}$ \Comment{Algorithm~\ref{algo:shifts}}}
    \For {$i = 1,\,\ldots,\,n_\gamma$ \textbf{parallel}} \label{line:par-spread-v}
        \Statex \Comment{Loop over IB points}
        \State $p \gets P_i$
        \State $\x \gets h(\idx{\X_p}+\stag)$\Comment{$\X_p\in\interface_h$, $\x\in\domain_h$}
        \State $\Delta\x \gets \x-\X_i$
        \State $w \gets \Dirac_h(\Delta\x+h\shift)$
        \State $V_i \gets w \cdot L_p$
    \EndFor \label{line:par-spread-vend}
    \State \textbf{reduce} $\{V_i\}$ \textbf{by} $\{K'_i\}$ \label{line:par-spread-reduce} \Comment{Segmented reduce}
    \For {$i = 1,\,\ldots,\,q$ \textbf{parallel}} \label{line:par-spread-quse}
        \Statex\Comment{Loop over inhabited grid cells, $q \le n_\gamma$}
        \State $\x \gets h(\key^{-1}(K'_i) + \stag)$
        \State $m \gets \#(\idx{\x} + \shift)$ \Comment{$\idx{\x}\equiv\key^{-1}(K'_i)$}
        \If {$m\neq\error$}
            \State $\ell_m \gets \ell_m + V_i$
        \EndIf
    \EndFor \label{line:par-spread-wend}
\EndFor
\State \Return $\vec{\ell}$
\EndProcedure
\end{algorithmic}
\end{algorithm}

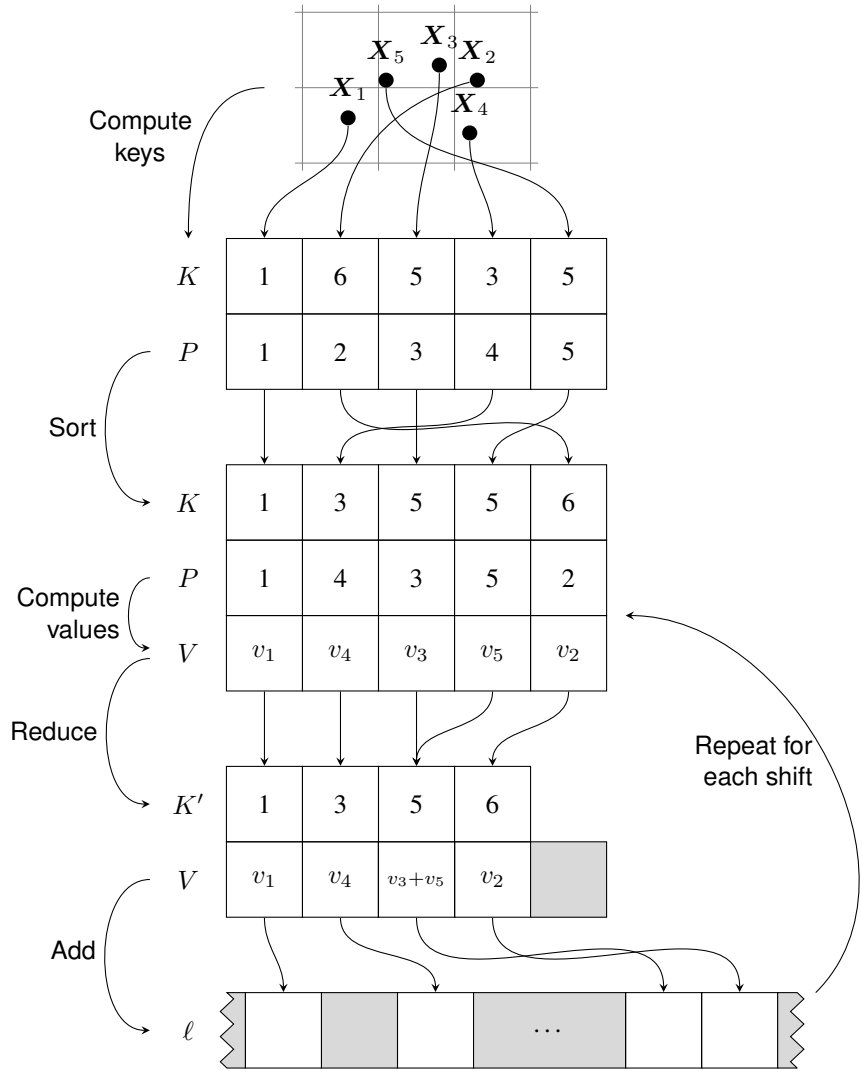
\begin{figure*}[ht]
    \begin{center}
    \begin{tikzpicture}
        \draw[help lines] (-0.1, -0.1) grid (3.1, 2.1);
        \node[circle, fill=black, inner sep=2pt, label={$\X_1$}] (x1) at (0.6, 0.6) {};
        \node[circle, fill=black, inner sep=2pt, label={$\X_5$}] (x2) at (1.1, 1.1) {};
        \node[circle, fill=black, inner sep=2pt, label={$\X_3$}] (x3) at (1.8, 1.3) {};
        \node[circle, fill=black, inner sep=2pt, label={$\X_2$}] (x4) at (2.3, 1.1) {};
        \node[circle, fill=black, inner sep=2pt, label={$\X_4$}] (x5) at (2.2, 0.4) {};

        \node[draw=none, rectangle, minimum width=1cm, minimum height=1cm, text centered] (kh) at (-1.5, -1.5) {$K$};
        \node[draw, rectangle, minimum width=1cm, minimum height=1cm, right of=kh] (k1) {1};
        \node[draw, rectangle, minimum width=1cm, minimum height=1cm, right of=k1] (k2) {6};
        \node[draw, rectangle, minimum width=1cm, minimum height=1cm, right of=k2] (k3) {5};
        \node[draw, rectangle, minimum width=1cm, minimum height=1cm, right of=k3] (k4) {3};
        \node[draw, rectangle, minimum width=1cm, minimum height=1cm, right of=k4] (k5) {5};

        \draw[->, >=stealth] (-0.5, 1) to [out=180, in=90] node [midway, left] {\sffamily\shortstack{Compute\\keys}} (-1.5, -1);

        \node[draw=none, rectangle, minimum width=1cm, minimum height=1cm, text centered, below of=kh] (ph) {$P$};
        \node[draw, rectangle, minimum width=1cm, minimum height=1cm, right of=ph] (p1) {1};
        \node[draw, rectangle, minimum width=1cm, minimum height=1cm, right of=p1] (p2) {2};
        \node[draw, rectangle, minimum width=1cm, minimum height=1cm, right of=p2] (p3) {3};
        \node[draw, rectangle, minimum width=1cm, minimum height=1cm, right of=p3] (p4) {4};
        \node[draw, rectangle, minimum width=1cm, minimum height=1cm, right of=p4] (p5) {5};

        \draw[->, >=stealth] (x1) to [out=270, in=90] (k1.north);
        \draw[->, >=stealth] (x4) to [out=195, in=90] (k2.north);
        \draw[->, >=stealth] (x3) to [out=270, in=90] (k3.north);
        \draw[->, >=stealth] (x5) to [out=270, in=90] (k4.north);
        \draw[->, >=stealth] (x2) to [out=270, in=90] (k5.north);

        \node[draw=none, rectangle, minimum width=1cm, minimum height=1cm, text centered] (lh) at (-1.5, -4.5) {$K$};
        \node[draw, rectangle, minimum width=1cm, minimum height=1cm, right of=lh] (l1) {1};
        \node[draw, rectangle, minimum width=1cm, minimum height=1cm, right of=l1] (l2) {3};
        \node[draw, rectangle, minimum width=1cm, minimum height=1cm, right of=l2] (l3) {5};
        \node[draw, rectangle, minimum width=1cm, minimum height=1cm, right of=l3] (l4) {5};
        \node[draw, rectangle, minimum width=1cm, minimum height=1cm, right of=l4] (l5) {6};

        \draw[->, >=stealth] (-2, -2.5) to [out=180, in=180] node [midway, left] {\sffamily Sort} (-2, -4.5);

        \node[draw=none, rectangle, minimum width=1cm, minimum height=1cm, text centered, below of=lh] (qh) {$P$};
        \node[draw, rectangle, minimum width=1cm, minimum height=1cm, right of=qh] (q1) {1};
        \node[draw, rectangle, minimum width=1cm, minimum height=1cm, right of=q1] (q2) {4};
        \node[draw, rectangle, minimum width=1cm, minimum height=1cm, right of=q2] (q3) {3};
        \node[draw, rectangle, minimum width=1cm, minimum height=1cm, right of=q3] (q4) {5};
        \node[draw, rectangle, minimum width=1cm, minimum height=1cm, right of=q4] (q5) {2};

        \draw[->, >=stealth] (p1) to [out=270, in=90] (l1.north);
        \draw[->, >=stealth] (p2) to [out=270, in=90] (l5.north);
        \draw[->, >=stealth] (p3) to [out=270, in=90] (l3.north);
        \draw[->, >=stealth] (p4) to [out=270, in=90] (l2.north);
        \draw[->, >=stealth] (p5) to [out=270, in=90] (l4.north);

        \node[draw=none, rectangle, minimum width=1cm, minimum height=1cm, text centered, below of=qh] (vh) {$V$};
        \node[draw, rectangle, minimum width=1cm, minimum height=1cm, right of=vh] (v1) {$v_1$};
        \node[draw, rectangle, minimum width=1cm, minimum height=1cm, right of=v1] (v2) {$v_4$};
        \node[draw, rectangle, minimum width=1cm, minimum height=1cm, right of=v2] (v3) {$v_3$};
        \node[draw, rectangle, minimum width=1cm, minimum height=1cm, right of=v3] (v4) {$v_5$};
        \node[draw, rectangle, minimum width=1cm, minimum height=1cm, right of=v4] (v5) {$v_2$};

        \draw[->, >=stealth] (qh.west) to [out=180, in=180] node[midway, left] {\sffamily\shortstack[r]{Compute\\values}} ($(vh.west) + (0, 2pt)$);

        \node[draw=none, rectangle, minimum width=1cm, minimum height=1cm, text centered] (mh) at (-1.5, -8.5) {$K'$};
        \node[draw, rectangle, minimum width=1cm, minimum height=1cm, right of=mh] (m1) {1};
        \node[draw, rectangle, minimum width=1cm, minimum height=1cm, right of=m1] (m2) {3};
        \node[draw, rectangle, minimum width=1cm, minimum height=1cm, right of=m2] (m3) {5};
        \node[draw, rectangle, minimum width=1cm, minimum height=1cm, right of=m3] (m4) {6};

        \draw[->, >=stealth] ($(vh.west)+(0, -2pt)$) to [out=180, in=180] node [midway, left] {\sffamily Reduce} (mh.west); 
        \node[draw=none, rectangle, minimum width=1cm, minimum height=1cm, text centered, below of=mh] (wh) {$V$};
        \node[draw, rectangle, minimum width=1cm, minimum height=1cm, right of=wh] (w1) {$v_1$};
        \node[draw, rectangle, minimum width=1cm, minimum height=1cm, right of=w1] (w2) {$v_4$};
        \node[draw, rectangle, minimum width=1cm, minimum height=1cm, right of=w2] (w3) {$\scriptstyle v_3+v_5$};
        \node[draw, rectangle, minimum width=1cm, minimum height=1cm, right of=w3] (w4) {$v_2$};
        \node[draw, rectangle, fill=black!15, minimum width=1cm, minimum height=1cm, right of=w4] {};

        \draw[->, >=stealth] (v1) to [out=270, in=90] (m1.north);
        \draw[->, >=stealth] (v2) to [out=270, in=90] (m2.north);
        \draw[->, >=stealth] (v3) to [out=270, in=90] (m3.north);
        \draw[->, >=stealth] (v4) to [out=270, in=90] (m3.north);
        \draw[->, >=stealth] (v5) to [out=270, in=90] (m4.north);

        \node[draw=none, rectangle, minimum width=1cm, minimum height=1cm, text centered] (fh) at (-1.5, -11.5) {$\ell$};
        \draw[fill=black!15] ($(fh.north east) + (0.25cm-0.2pt, -0.2pt)$) -- ($(fh.north east) + (0cm, -0.2pt)$) to [snake=zigzag, segment length=0.33333205cm] ($(fh.south east) + (0cm, 0.2pt)$) -- ($(fh.south east) + (0.25cm-0.2pt, 0.2pt)$) -- cycle;
        \node[draw, rectangle, minimum width=1cm, minimum height=1cm, text centered, right of=fh, xshift=0.25cm] (f1) {};
        \node[draw, rectangle, fill=black!15, minimum width=1cm, minimum height=1cm, text centered, right of=f1] (f2) {};
        \node[draw, rectangle, minimum width=1cm, minimum height=1cm, text centered, right of=f2] (f3) {};
        \node[draw, rectangle, fill=black!15, minimum width=2cm, minimum height=1cm, text centered, right of=f3, xshift=0.5cm] (fe) {\dots};
        \node[draw, rectangle, minimum width=1cm, minimum height=1cm, text centered, right of=fe, xshift=0.5cm] (f4) {};
        \node[draw, rectangle, minimum width=1cm, minimum height=1cm, text centered, right of=f4] (f5) {};
        \draw[fill=black!15] ($(f5.north east) + (-0.2pt, -0.2pt)$) -- ($(f5.north east) + (0.25cm, -0.2pt)$) to [snake=zigzag, segment length=0.33333205cm] ($(f5.south east) + (0.25cm, 0.2pt)$) -- ($(f5.south east) + (-0.2pt, 0.2pt)$) -- cycle;

        \draw[->, >=stealth] (wh.west) to [out=180, in=180] node [midway, left, yshift=2pt] {\sffamily Add} (fh.west);

        \draw[->, >=stealth] (w1.south) to [out=270, in=90] (f1.north);
        \draw[->, >=stealth] (w2.south) to [out=270, in=90] (f3.north);
        \draw[->, >=stealth] (w3.south) to [out=270, in=90] (f4.north);
        \draw[->, >=stealth] (w4.south) to [out=270, in=90] (f5.north);

        \draw[->, >=stealth] (6.75, -11) to [out=45, in=0] node [midway, left, xshift=-2pt] {\sffamily\shortstack[r]{Repeat for\\each shift}} (4.25, -6);
    \end{tikzpicture}
    \end{center}

    \caption{%
Schematic of Algorithm~\ref{algo:par-spread} for Lagrangian points $\X_1$
through $\X_5$ in a small region of $\domain$. Gray grid lines indicate the
boundaries of grid cells. The rows labelled $K$, $P$, $V$, and $\ell$ represent
the memory of arrays for the keys, permutation, values, and output vector,
respectively. Keys are computed according to grid cell, starting with 1 at the
bottom left and continuing sequentially from left to right then bottom to top.
The value $v_i$ corresponds to point $\X_i$ for a given shift, and are computed
in sorted order using the values in $P$. The results of reducing $V$ by $K$
yields unique keys $K'$. Points $\X_3$ and $\X_5$ inhabit the same grid cell,
so reduction adds their corresponding values before writing to the output
vector. Gray regions denote unused memory.
    }\label{fig:algo}
\end{figure*}

Lastly, we need suitable way to generate keys. A function $\key$ that generates keys
should be 1-to-1 with grid cells, or alternatively, to points in $\bar{\domain}_h$. For
this reason, it is useful to formulate $\key$ as a function of $\mathbb{Z}^d$ and so
$\key(\idx{\X})$ gives the key for IB point $\X$. The requirement that $\key$ be
injective will, in general, invalidate the grid indexing function as an otherwise good
choice, as it maps each ghost point in $\bar{\domain}_h$ to $\error$. We can, however,
use the key to compute grid indices: the point in $\domain_h$ with key $k$ also has grid
index $\#(\key^{-1}(k))$. Because $\key$ is injective, $\key^{-1}(k)$ is well-defined,
and for shift $\shift$, $\#(\key^{-1}(k)+\shift)$ yields the appropriate target index for
writing to the output vector. Putting these pieces together, we have a complete parallel
spreading algorithm, listed in Algorithm \ref{algo:par-spread} and illustrated in Figure~
\ref{fig:algo}.

Lines \ref{line:par-spread-ps}--\ref{line:par-spread-sort} of Algorithm
\ref{algo:par-spread} construct the permutation by constructing a list of keys and a list
of the indices of IB points, $1,\,2,\,\ldots,\,n_\gamma$, and sorting the indices
according to the keys. The $i\th$ entry in the permuted list of indices gives the index
of the $i\th$ IB point in sorted order. On line \ref{line:par-spread-q}, we define $q$
to be the number of unique keys, which is also the number of reduced values to write.
Since the list of sorted keys does not change for different shifts, we compute it once
and reuse the value; see lines \ref{line:par-spread-quse}--\ref{line:par-spread-wend}.
Values, one per IB point, are computed and stored in a list $\vec{V}$ on lines
\ref{line:par-spread-v}--\ref{line:par-spread-vend}. These values are then input to the
segmented reduce, line \ref{line:par-spread-reduce}.

Algorithm \ref{algo:par-spread} does not explicitly depend on the Eulerian grid, with
some caveats dependent upon implementation details. Our implementation relies on the
{\thrust} library to provide the key-value sort, segmented reduce, and unique counting
routines. Sorting is implemented as radix sort, which has a runtime of
$\bigo{wn_\gamma/p}$, where $p$ is the number of processors/threads, and $w$ is the
number of bits required to represent every key. In general, $w\propto\log_2 n$, for $n$
elements to be sorted, but we use 32-bit integers for keys, so $w=32$. It is reasonable
to assume that this is true for most use-cases, as there are approximately $n_\omega$
possible keys, and implementations of \texttt{BLAS} and \texttt{LAPACK} typically use
32-bit integers for indexing. However, extremely fine grids with more than $2^{32}$ grid
points will require a larger data type to represent each key uniquely. In that case, $w$
increases with a finer grid. Segmented reduction has a much more complicated relationship
with the Eulerian grid. Parallelized segmented reduction has a worst-case runtime of
$\bigo{n_\gamma/p}$, but the constant of proportionality depends on the density of IB
points within inhabited grid cells. On the one hand, if all IB points inhabit the same
grid cell, segmented reduce proceeds as a regular reduce, which is very fast. On the
other hand, if every IB point inhabits its own grid cell, there is no work to be done,
and any time spent by the algorithm is for naught. The density of IB points also affects
the value of $q$, but is bounded above by $n_\gamma$. For up to $2^{32}$ grid points, we
expect the overall runtime of the spreading algorithm to be $\bigo{n_\gamma/p}$.

Finally, we remark that Algorithm \ref{algo:par-spread} must synchronize threads once per
shift. For choices of $\kernel$ where $s^d$ is large, this can hurt performance. This
requirement is written as a parallel loop (line \ref{line:par-spread-v}) within a serial
loop (line \ref{line:par-spread-shifts}). To reduce the number of synchronizations, we
must be able to compute and store several values at once. However, attempting to write
multiple values of once may lead to contentions. Below, we develop algorithms that
require fewer synchronizations by using more memory to spread several values at once and
avoid write contentions.

\subsection{Buffered spreading variants}

Here we introduce a buffer in which to store incomplete calculations. It can be thought
of as a set of temporary output vectors, which we will combine at the end of the
algorithm to finish the calculation. These vectors have $n_\omega$ entries, so adding
them requires as many operations, but is quite easily parallelized, and is provided by
any parallel \texttt{BLAS} implementation. In this regard, these buffered variants depend
explicitly on the Eulerian grid. The only additional requirement is that there is enough
memory to hold the buffer. 

To develop these variants, we note that the sorted list of keys and the permutation
are the same for all shifts. Since the permutation dictates the order of the values to
spread, and the list of keys decides the behavior of the segmented reduce, these steps
are nearly identical for each shift, save for the effect each shift might have on the
values. We can therefore use several shifts to compute one value of $\Dirac_h$ each, and
use the same machinery to accumulate the values as if they were vectors. Then, to avoid
write contentions, each of the entries of the reduced vectors is written to a separate
output vector of the buffer. We choose $\bufsz\approx 10$ values to compute each
iteration. Summing the vectors in the buffer yields the desired result.

\begin{algorithm}
\caption{Pre-allocated buffer parallel spreading}
\label{algo:pa-spread}
\begin{algorithmic}[1]
\Procedure{pa-par-spread}{$\interface_h,\,\domain_h,\,\vec{L}, \vec{\ell}_1,\,\ldots,\,\vec{\ell}_\bufsz$}
\State $\triangleright\ $\textbf{require}: $\bufsz \ge 1$
\State $\triangleright\ $\textbf{generate}: {\revb\parbox[t]{\dimexpr\linewidth-1.5em-\genlinewidth}{Approximate values of $\ell$ (Eq.~\eqref{eq:scalar-spread}) at each point in $\domain_h$}}
\For {$i = 1,\,\ldots,\,n_\gamma$ \textbf{parallel}}
    \Statex \Comment{Loop over IB points}
    \State $K_i \gets \key(\idx{\X_i})$ \Comment{Sort key}
    \State $P_i \gets i$ \Comment{Initial ordering}
\EndFor
\State \textbf{sort} $\{P_i\}$ \textbf{by} $\{K_i\}$ \Comment{$i\to P_i$ is a permutation}
\State $q \gets \text{\textbf{count unique} }\{K_i\}$
\For {$j = {\revb0},\,\ldots,\,\ceil{s^d/\bufsz}{\revb-1}$} \Comment{Loop over shifts}
    \State $\{K'_i\} \gets \{K_i\}$
    \For {$i = 1,\,\ldots,\,n_\gamma$ \textbf{parallel}}
        \Statex \Comment{Loop over IB points}
        \State $p \gets P_i$
        \State $\x \gets h(\idx{\X}+\stag)$\Comment{$\X_p\in\interface_h$, $\x\in\domain_h$}
        \State $\Delta\x \gets \x-\X_i$
        \For {$k=1,\,\ldots,\,\textbf{min}(\bufsz,\,s^d-\bufsz\cdot j)$}
            \State {\revb$\shift \gets \Call{shift}{\bufsz\cdot j + k,\,s,\,d}$}
            \Statex {\revb\Comment{Algorithm~\ref{algo:shifts}}}
            \State $w \gets \Dirac_h(\Delta\x+h\shift)$
            \State $V_{ik} \gets w \cdot L(\X_p)$ \Comment{$V\in\mathbb{R}^{n_\gamma\times\bufsz}$}
        \EndFor
    \EndFor
    \State \textbf{reduce} $\{V_{i\cdot}\}$ \textbf{by} $\{K'_i\}$
    \For {$i = 1,\,\ldots,\,q$ \textbf{parallel}}
        \Statex \Comment{Loop over inhabited grid cells, $q\le n_\gamma$}
        \State $\x \gets h(\key^{-1}(K'_i) + \stag)$
        \For {$k=1,\,\ldots,\,\textbf{min}(\bufsz,\,s^d-\bufsz\cdot j)$}
            \State {\revb$\shift \gets \Call{shift}{\bufsz\cdot j + k,\,s,\,d}$}
            \Statex {\revb\Comment{Algorithm~\ref{algo:shifts}}}
            \State $m \gets \#(\idx{\x} + \shift)$
            \If {$m\neq\error$}
                \State $\ell_{mk} \gets \ell_{mk} + V_{ik}$
            \EndIf
        \EndFor
    \EndFor
\EndFor
\State \Return {\revb$\sum_{k=1}^{\bufsz} \vec{\ell}_k$}
\EndProcedure
\end{algorithmic}
\end{algorithm}

\begin{algorithm}
\caption{On-the-fly buffer parallel spreading}
\label{algo:otf-spread}
\begin{algorithmic}[1]
\Procedure{otf-par-spread}{$\interface_h,\,\domain_h,\,\vec{L}$}
\State $\triangleright\ $\textbf{require}: $\bufsz \ge 1$
\State $\triangleright\ $\textbf{generate}: {\revb\parbox[t]{\dimexpr\linewidth-1.5em-\genlinewidth}{Approximate values of $\ell$ (Eq.~\eqref{eq:scalar-spread}) at each point in $\domain_h$}}
\For {$i = 1,\,\ldots,\,\bufsz$}
    \State $\vec{\ell}_k \gets \vec{0}$
\EndFor
\State \Return \Call{pa-par-spread}{$\interface_h,\,\domain_h,\,\vec{L},\,\vec{\ell}_1,\,\ldots,\,\vec{\ell}_\bufsz$}
\Statex \Comment{Algorithm \ref{algo:pa-spread}}
\EndProcedure \Comment{Lifetime of $\vec{\ell}_k$ ends here}
\end{algorithmic}
\end{algorithm}

Algorithm \ref{algo:pa-spread} lists the general form of the algorithm as pseudocode, and
Algorithm \ref{algo:otf-spread} gives a minor modification. The difference between these
two variants is the lifetime of the buffer: Algorithm \ref{algo:pa-spread} does not
manage the buffer itself, but Algorithm \ref{algo:otf-spread} allocates and frees the
buffer memory, limiting the lifetime of the buffer to the duration of the algorithm. The
latter considers the memory allocation as part of the algorithm. The additional overhead
from buffer allocation means that we never expect Algorithm \ref{algo:otf-spread} to
outperform Algorithm \ref{algo:pa-spread}; it is provided as an alternative for systems
where maintaining a large, long-lived buffer may exhaust memory. We compare these
algorithms and Algorithm \ref{algo:par-spread} to Algorithm \ref{algo:serial-spread}, and
the efficacy of Algorithm \ref{algo:par-interp} below.



\section{Numerical results} \label{sec:results}

Here we describe two types of test: unstructured IB points, in which points are placed
randomly in the domain and generate a force independently from the other IB points, and
structured IB points, in which the points comprise an elastic structure and forces are
generated based on the configuration of the points as a whole. For these tests, we use
a $16\um\times16\um\times16\um$ triply periodic domain with an initially shear-like flow,
$\u=(0,\,0,\,\shear(y-8\um))$, with shear rate $\shear$. Tests use a shear rate of
$1000\si{\per\second}$ unless otherwise noted. This flow has a sharp transition at the
periodic boundary $y=0\um$, so a background force is added to maintain this transition
and so that the initial flow is also the steady flow in the absence of other forces.

Serial and multicore CPU tests were performed on a single node with 48 Intel{\reg}
Xeon{\reg} CPU E5-2697 v2 2.70\si{\giga\hertz} processors and 256 GB of RAM running
CentOS Linux release 7.7.1908 (x86\_64). Parallel CPU implementations use Intel's OpenMP
library, \texttt{libiomp5}. GPU tests used the same node with an NVIDIA{\reg} Tesla{\reg}
K80 ($2\times$GK210 GPU with 13 823.5\si{\mega\hertz} multiprocessors and 12 GB of global
memory each). Only one of the GK210 GPUs was used. The CPU code was written in C++20 and
the GPU code was written in C++/CUDA and used version 9.2 CUDA libraries (\cite{ibcu}).
Both the CPU and GPU code were compiled using \texttt{clang} version 7.0.1. All tests are
performed in double precision. We begin with tests using unstructured IB points, for
which both of these architectures were used.

\subsection{Unstructured IB points}\label{sec:unst}

Consider a set of $n$ IB points randomly placed in the domain described above. The IB
points are imagined to be tethered to their initial positions. The fluid solver is not
invoked for these tests. Instead, at each timestep, we interpolate the fluid velocity to
each of the IB points and predict updated positions for the IB points. Using these
predicted positions, we compute a Hookean force for each IB point with spring constant
$0.01\si{dyn\per\centi\meter}$. We spread these forces from the predicted positions to
the fluid grid, but do not use them to update the fluid velocity. This ensures that the
points do not settle into a steady position so the spreading and interpolation operations
receive new data each timestep. Finally, we interpolate the velocity to the positions of
the IB points at the beginning of the timestep again and update the position of the IB
points. While the interpolated velocities are the same as those computed at the beginning
of the timestep, this is done by analogy to the fluid solver, which interpolates fluid
velocities twice and spreads forces once per timestep.

We use this test to compare the performance of the parallel algorithms to their serial
counterparts and, for the spreading variants, to each other.

\subsubsection{Dependence on background grid}\label{sec:grid-dependence}

The serial Algorithms~\ref{algo:serial-spread} and~\ref{algo:par-interp} do not
explicitly depend on the size of the fluid grid. With perhaps the exception of the
sorting and reducing steps, Algorithm~\ref{algo:par-spread} also does not depend on the
size of the grid. Algorithms~\ref{algo:pa-spread} and~\ref{algo:otf-spread}, on the other
hand, ultimately sum their buffer vectors, which have one entry per grid point.
Algorithm~\ref{algo:otf-spread} also incorporates the allocation of these buffers. Using
a few different fluid grids, we investigate whether Algorithms~\ref{algo:par-interp} and%
~\ref{algo:par-spread} are independent of the grid in practice, and how the grid
dependence affects the runtime of Algorithms~\ref{algo:pa-spread} and~%
\ref{algo:otf-spread}.

\begin{table*}[ht]
    \caption{%
Average timing results for interpolating to and spreading from $2^{16}$ IB points from
1000 timesteps on different devices and grid configurations. Interpolation (Algorithm~%
\ref{algo:par-interp}) happens twice per timestep and spreading (Algorithms~%
\ref{algo:serial-spread},~\ref{algo:par-spread}--\ref{algo:otf-spread}) happens once per
timestep. Grid refinement is the number of grid points per $16\um$ in each dimension.
Times per call to the listed algorithm are reported in seconds.
    }\label{tab:grid-timing}
    \begin{center}
    \bgroup
    \renewcommand{\arraystretch}{1.7}
    \begin{tabular}{ccccccc}
                                                                                              \toprule
                             &                          & \multicolumn{4}{c}{Grid refinement}   \\ \cline{3-6}
        Device               & Algorithm                & 16      & 32      & 64      & 128     \\ \midrule
        $1\times\text{CPU}$  & \revb\ref{algo:serial-spread} & \revb1.33249 & \revb1.33621 & \revb1.33840 & \revb1.37281 \\
                             & \revb\ref{algo:par-interp}    & \revb1.29633 & \revb1.31373 & \revb1.30763 & \revb1.35101 \\ \midrule
        $16\times\text{CPU}$ & \ref{algo:par-interp}    & 0.09890 & 0.09928 & 0.09974 & 0.10624 \\
                             & \ref{algo:par-spread}    & 0.23282 & 0.26431 & 0.25783 & 0.26590 \\
                             & \ref{algo:pa-spread}     & 0.12803 & 0.14213 & 0.15215 & 0.20107 \\
                             & \ref{algo:otf-spread}    & 0.12965 & 0.14242 & 0.14766 & 0.21874 \\ \midrule
        $1\times\text{GPU}$  & \ref{algo:par-interp}    & 0.01253 & 0.01317 & 0.01722 & 0.01816 \\
                             & \ref{algo:par-spread}    & 0.03930 & 0.04020 & 0.04215 & 0.04755 \\
                             & \ref{algo:pa-spread}     & 0.01715 & 0.02049 & 0.02370 & 0.03656 \\
                             & \ref{algo:otf-spread}    & 0.01804 & 0.02198 & 0.02873 & 0.07288 \\ \bottomrule
    \end{tabular}
    \egroup
    \end{center}
\end{table*}

Table \ref{tab:grid-timing} lists timing results for $n=2^{16}$ IB points and 16, 32, 64,
and 128 grid points per $16\um$. The rows with device listed as $1\times\text{CPU}$
correspond to the serial algorithms and will serve as a reference point for the rest of
the section. If the serial algorithms depend on the fluid grid, they do so only mildly.
In fact, under close scrutiny, it seems that these deviations are due to hardware-level
differences in the integer multiplications and additions used in computing sort keys and
grid indices. We can therefore expect each of the algorithms to exhibit a mild variation
in runtime for different grid refinements.

\begin{figure}[t]
\begin{tikzpicture}
\begin{groupplot}[
    group style={group name=dep, group size=1 by 2},
    width=0.45\textwidth
]
\nextgroupplot[
        xmin=12,
        xmax=160,
        xmode=log,
        log basis x=2,
        ymin=3,
        ymax=24,
        ymode=log,
        log basis y=2,
        log origin=infty,
        ytick={2, 4, 8, 16},
        width=0.45\textwidth,
        height=0.45\textwidth,
        axis lines=center,
        ylabel={speedup},
        xlabel near ticks,
        ylabel near ticks
    ]
    
    \addplot+[only marks, mark=diamond*, color=tol/vibrant/magenta, mark options={scale=2, fill=tol/vibrant/magenta}] coordinates {%
        (16 , {1.44570/0.09890})
        (32 , {1.45665/0.09928})
        (64 , {1.46783/0.09974})
        (128, {1.51988/0.10624})
    }; \label{plot:grid-dep-interp};
    \addplot+[only marks, mark=square*, color=tol/vibrant/teal, mark options={scale=2, fill=tol/vibrant/teal}] coordinates {%
        (16 , {1.47196/0.23282})
        (32 , {1.48014/0.26431})
        (64 , {1.48519/0.25783})
        (128, {1.53801/0.26590})
    }; \label{plot:grid-dep-spread};
    \addplot+[only marks, mark=*, color=tol/vibrant/orange, mark options={scale=2, fill=tol/vibrant/orange}] coordinates {%
        (16 , {1.47196/0.12803})
        (32 , {1.48014/0.14213})
        (64 , {1.48519/0.15215})
        (128, {1.53801/0.20107})
    }; \label{plot:grid-dep-pa};
    \addplot+[only marks, mark=triangle*, color=tol/vibrant/blue, mark options={scale=2, fill=tol/vibrant/blue}] coordinates {%
        (16 , {1.47196/0.12965})
        (32 , {1.48014/0.14242})
        (64 , {1.48519/0.14766})
        (128, {1.53801/0.21874})
    }; \label{plot:grid-dep-otf};
    \node [fill=white] at (rel axis cs: 0.075, 0.95) {\sffamily(a)};
\nextgroupplot[
    xmin=12,
    xmax=192,
    xmode=log,
    log basis x=2,
    ymin=12,
    ymax=160,
    ymode=log,
    log basis y=2,
    log origin=infty,
    ytick={16, 32, 64, 128},
    width=0.45\textwidth,
    height=0.45\textwidth,
    axis lines=center,
    xlabel={grid refinement},
    ylabel={speedup},
    xlabel near ticks,
    ylabel near ticks
]
    \addplot+[only marks, mark=diamond*, color=tol/vibrant/magenta, mark options={scale=2, fill=tol/vibrant/magenta}] coordinates {%
        (16 , {1.44570/0.01253})
        (32 , {1.45665/0.01317})
        (64 , {1.46783/0.01722})
        (128, {1.51988/0.01816})
    };
    \addplot+[only marks, mark=square*, color=tol/vibrant/teal, mark options={scale=2, fill=tol/vibrant/teal}] coordinates {%
        (16 , {1.47196/0.03930})
        (32 , {1.48014/0.04020})
        (64 , {1.48519/0.04215})
        (128, {1.53801/0.04755})
    };
    \addplot+[only marks, mark=*, color=tol/vibrant/orange, mark options={scale=2, fill=tol/vibrant/orange}] coordinates {%
        (16 , {1.47196/0.01715})
        (32 , {1.48014/0.02049})
        (64 , {1.48519/0.02370})
        (128, {1.53801/0.03656})
    };
    \addplot+[only marks, mark=triangle*, color=tol/vibrant/blue, mark options={scale=2, fill=tol/vibrant/blue}] coordinates {%
        (16 , {1.47196/0.01804})
        (32 , {1.48014/0.02198})
        (64 , {1.48519/0.02873})
        (128, {1.53801/0.07288})
    };
    \node [fill=white] at (rel axis cs: 0.075, 0.95) {\sffamily(b)};
\end{groupplot}
\path (dep c1r2.south west|-current bounding box.south)--
coordinate(legendpos) (dep c1r2.south east|-current bounding box.south);
\matrix[
    matrix of nodes,
    anchor=north,
    inner sep=0.2em,
  ]at([yshift=-1ex]legendpos)
  {
    \ref{plot:grid-dep-interp}& Algorithm \ref{algo:par-interp}&[5pt]
    \ref{plot:grid-dep-spread}& Algorithm \ref{algo:par-spread}&[5pt] \\
    \ref{plot:grid-dep-pa}& Algorithm \ref{algo:pa-spread}&[5pt]
    \ref{plot:grid-dep-otf}& Algorithm \ref{algo:otf-spread}\\};
\end{tikzpicture}
\caption{%
Speedup of Algorithm \ref{algo:par-interp} (interpolation) and Algorithms
\ref{algo:par-spread}--\ref{algo:otf-spread} (spreading) compared to their serial
counterparts on $2^{16}$ randomly placed IB points for different grid refinements on (a)
16 CPUs and (b) the GPU. 
}
\label{fig:grid-dependence}
\end{figure}
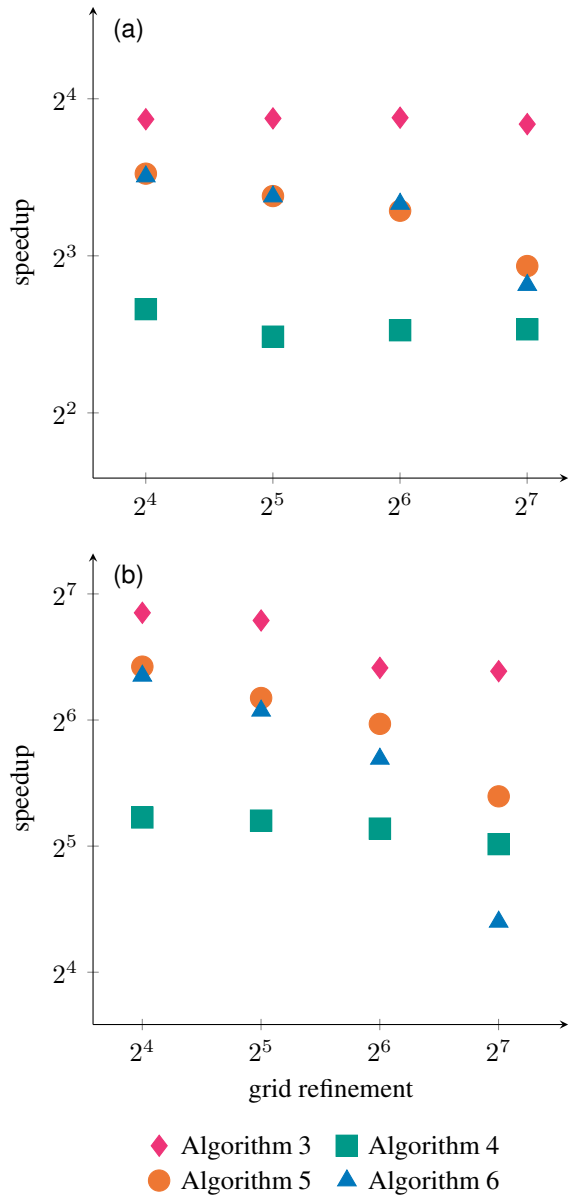

The speedup for these schemes is illustrated in Figure \ref{fig:grid-dependence}.
Algorithm \ref{algo:par-interp} is independent of the fluid grid, as expected. Any grid
dependence introduced by the sort and reduce steps of Algorithm \ref{algo:par-spread} is
not obvious for the grids presented. On the other hand, the degradation of speedup for
the sweep-fused Algorithms \ref{algo:pa-spread} and \ref{algo:otf-spread} is apparent for
finer grids. For small problems with $128^3$ or fewer grid points, one gets better
performance from the sweep-fused variants, with the exception of Algorithm
\ref{algo:otf-spread} on the GPU. This can be attributed to the slower allocation of the
buffer for a finer grid. For problems where the grid has $256^3$ or more grid points, we
expect Algorithm \ref{algo:par-spread} to be the fastest choice for spreading. Because
our fluid solver fits in GPU memory only for fewer than $128^3$ grid points, for the
remainder of this work we consider only a grid with a grid refinement of 64 ($h =
0.25\um$). We can imagine using this algorithm on a less capable device with more limited
memory, so we restrict ourselves to Algorithm \ref{algo:otf-spread} for spreading,
computing $\bufsz=8$ values per sweep, to minimize the lifetime of the buffer used in
spreading while still enjoying the benefit of using the buffer.


\subsubsection{Strong scaling}\label{sec:unst-strong}

It is commonly the case that one wishes to employ parallelization to improve runtimes for
a problem of interest. To illustrate this improvement, we now consider how runtime varies
for the test problem with $n=2^{16}$ IB points, a grid refinement of 64 (grid size
$h=0.25\um$), and Algorithm \ref{algo:otf-spread} with different numbers of threads. We
use up to 32 threads on the CPU and 64--4096 threads on the GPU. For a fixed problem, we
ideally wish to see the runtime using $2p$ threads to be half of that using $p$ threads.
In other words, using twice as many threads yield an ideal speedup of 2.

\begin{figure}[htbp]
\begin{tikzpicture}
\begin{groupplot}[
    group style={group name=unst-strong, group size=1 by 2},
    width=0.45\textwidth
]
\nextgroupplot[
        xmin=0.70710678118,  
        xmax=45.2548339959,  
        xmode=log,
        log basis x=2,
        ymode=log,
        ymin=0.5,
        ymax=128,
        log basis y=2,
        log origin=infty,
        width=0.45\textwidth,
        height=0.45\textwidth,
        axis lines=center,
        ylabel={speedup},
        xlabel near ticks,
        ylabel near ticks,
        xtick={1, 2, 4, 8, 16, 32},
        ytick={1, 4, 16, 64},
        legend style={at={(0.5, 0.9)}, anchor=center, draw=none,
                      /tikz/every even column/.append style={column sep=5pt}},
        legend cell align={left},
        legend columns=1
    ]
    \addplot+[only marks, mark=diamond*, color=tol/vibrant/magenta, mark options={scale=2, fill=tol/vibrant/magenta}] coordinates {%
        (1 , {1.30763/1.29442})
        (2 , {1.30763/0.65481})
        (4 , {1.30763/0.34712})
        (8 , {1.30763/0.18004})
        (16, {1.30763/0.09969})
        (32, {1.30763/0.07371})
    }; \label{plot:unst-strong-interp}
    \addplot+[only marks, mark=triangle*, color=tol/vibrant/blue, mark options={scale=2, fill=tol/vibrant/blue}] coordinates {%
        (1 , {1.33840/1.50377})
        (2 , {1.33840/0.92804})
        (4 , {1.33840/0.44697})
        (8 , {1.33840/0.24580})
        (16, {1.33840/0.14895})
        (32, {1.33840/0.12225})
    }; \label{plot:unst-strong-spread}
    \addplot+[no marks, dashed, black, domain=1:32] {x^0.93};
    \addplot+[no marks, dotted, black, domain=1:32] {0.88*x^0.89};
    \addplot+[no marks, black, domain=1:32] {x};
    \node[fill=white] at (rel axis cs: 0.075, 0.95) {\sffamily(a)};
\nextgroupplot[
        xmin=45.2548339959,  
        xmax=5792.61875148,  
        xmode=log,
        log basis x=2,
        ymode=log,
        ymin=0.5,
        ymax=128,
        log basis y=2,
        log origin=infty,
        width=0.45\textwidth,
        height=0.45\textwidth,
        axis lines=center,
        xlabel={threads},
        ylabel={speedup},
        xlabel near ticks,
        ylabel near ticks,
        ytick={1, 4, 16, 64},
        legend style={at={(0.5, 0.9)}, anchor=center, draw=none,
                      /tikz/every even column/.append style={column sep=5pt}},
        legend cell align={left},
        legend columns=1
    ]
    
    \addplot+[only marks, mark=diamond*, color=tol/vibrant/magenta, mark options={scale=2, fill=tol/vibrant/magenta}] coordinates {%
        (64  , {1.30763/0.80977})
        (128 , {1.30763/0.43929})
        (256 , {1.30763/0.22434})
        (512 , {1.30763/0.11258})
        (1024, {1.30763/0.05681})
        (2048, {1.30763/0.03074})
        (4096, {1.30763/0.01689})
    };
    \addplot+[only marks, mark=triangle*, color=tol/vibrant/blue, mark options={scale=2, fill=tol/vibrant/blue}] coordinates {%
        (64  , {1.33840/0.96698})
        (128 , {1.33840/0.49359})
        (256 , {1.33840/0.26208})
        (512 , {1.33840/0.14286})
        (1024, {1.33840/0.08252})
        (2048, {1.33840/0.05351})
        (4096, {1.33840/0.03905})
    };
    \addplot+[no marks, dashed, black, domain=64:4096] {1.61*(x/64)^0.93};
    \addplot+[no marks, dotted, black, domain=64:4096] {1.38*(x/64)^0.93};
    \addplot+[no marks, black, domain=64:4096] {1.61*x/64};
    \node[fill=white] at (rel axis cs: 0.075, 0.95) {\sffamily(b)};
\end{groupplot}
\path (unst-strong c1r2.south west|-current bounding box.south)--
coordinate(legendpos) (unst-strong c1r2.south east|-current bounding box.south);
\matrix[
    matrix of nodes,
    anchor=north,
    inner sep=0.2em,
  ]at([yshift=-1ex]legendpos)
  {
    \ref{plot:unst-strong-interp}& Algorithm \ref{algo:par-interp}&[5pt]
    \ref{plot:unst-strong-spread}& Algorithm \ref{algo:otf-spread}\\};
\end{tikzpicture}
\caption{%
Strong scaling results for Algorithm \ref{algo:otf-spread} and grid spacing
$h = 0.5\um$ (a grid refinement of 64) for $2^{16}$ randomly placed IB points
in a $16\um\times16\um\times16\um$ triply periodic domain for (a) 1-32 CPU
cores, and (b) 64-4096 threads on the GPU. Speedup is measured relative to
serial Algorithms \ref{algo:par-interp} and \ref{algo:serial-spread}. The solid
black lines show the trendline for ideal speedup. The dashed or dotted lines
give the initial trend for interpolation and spreading, respectively.
}
\label{fig:unstructured-strong}
\end{figure}
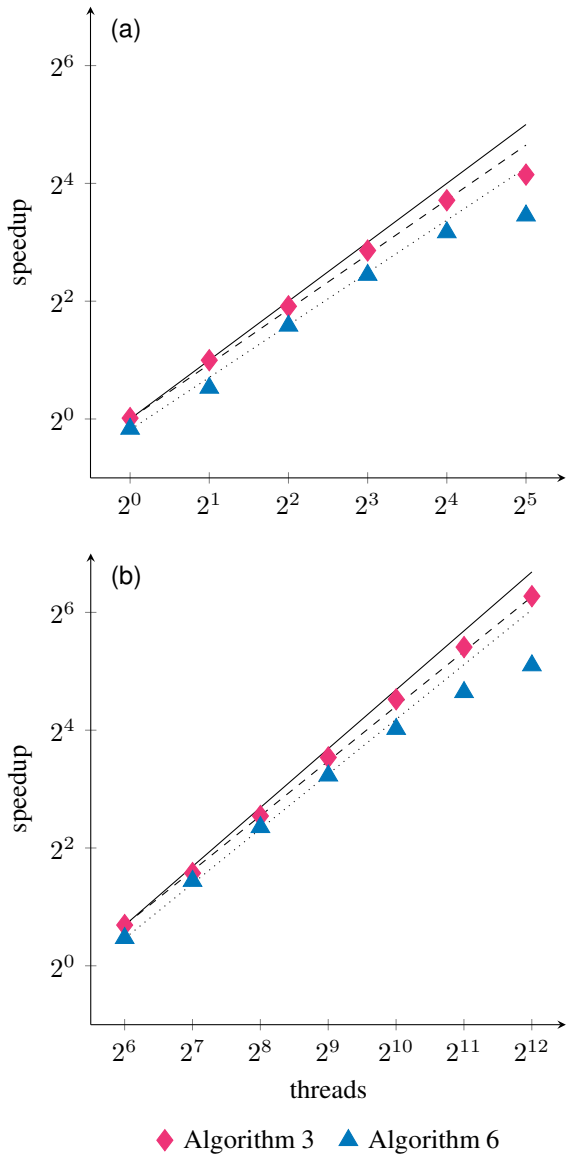

Figure \ref{fig:unstructured-strong} shows the results of these tests. Speedup is
measured relative to the serial interpolation and spreading implementations.  The
trendlines estimate that increasing computing resources by a factor of two decreases
runtime by a factor of about 1.91 for CPU and GPU interpolation and by a factor of about
1.85 for CPU spreading. It is not trivial to limit the number of threads used by
{\thrust} for work done on the GPU, so the key-value sort and segmented reduce use as
many threads as {\thrust} decides is prudent.  While the trendline indicates a decrease
in runtime by a factor of 1.91 as well, this is merely an approximation.  

Parallel CPU interpolation using a single processor is identical to the serial
CPU interpolate, so the CPU interpolate passes through $(1,\,1)$. The same is
not true of parallel spreading using a single processor compared to serial
spreading. Because of the additional sort step in the parallel spreading
algorithm, single-threaded Algorithm \ref{algo:par-spread} is about 12\% slower
than its serial counterpart. The CPU code also enjoys the benefit of using
vector registers for some of the computation. The GPU requires 64 threads to
match the speed of a single CPU core.
%
Even at 4096 threads, interpolate on the GPU shows no indication of plateauing. The final
data point for that curve shows a speedup of approximately $77\times$. Figure
\ref{fig:grid-dependence}(b), on the other hand, shows that the maximum speedup we can
expect for this problem is approximately 85$\times$, which the trendline in Figure
\ref{fig:unstructured-strong}(b) predicts will occur at approximately 4480 threads. Thus,
we can expect the plateau for interpolate on the GPU to be very abrupt. This indicates a
hardware limitation, the likely culprit being exhaustion of register memory. The
plateauing of the CPU curves is not a limitation of the algorithm for the CPU. Despite
having 48 cores, the test using 32 cores did not utilize them at full capacity. Using
fewer cores, on the other hand, was able to maintain full utilization for the duration of
the test. If not for having a comparatively limited number of CPU cores, we expect to see
the CPU trend continue.

If not for hardware limitations, it seems that the algorithm scales without bound on
either the CPU of GPU. Overall, trends for the CPU and GPU are very similar. Because of
these similarities, we will restrict ourselves to the GPU, but expect any conclusions to
hold for the CPU as well.

\subsubsection{Weak scaling}\label{sec:unst-weak}
In contrast, with improved computing resources, we may wish to solve bigger problems. The
ideal parallel algorithm solves a problem with $p$ threads in the same time as it solves
a twice bigger problem with $2p$ threads. Here, we place between $2^{16}$ and $2^{19}$
points randomly in the domain. We increase the number of threads proportionally, between
128 and 1024.  

\begin{table}[ht]
    \caption{%
Weak scaling results for interpolation and spreading for $p$ threads and $n$ randomly
placed IB points in a $16\um\times16\um\times16\um$ triply periodic domain with
$h=0.5\um$ on the GPU. Average time per call is reported in seconds. $N$ is the number of
samples taken.
    }\label{tab:unstructured-weak}
    \begin{center}
        \begingroup
        \setlength{\tabcolsep}{9pt}
        \renewcommand{\arraystretch}{1.5}
        \begin{tabular}{ccccc}
                                                                                               \toprule
            $p$  & $n$      & \titletable{interpolate}{20000} & \titletable{spread}{10000} \\ \midrule
            128  & $2^{16}$ & $0.43930$                       & $0.47632$                  \\
            256  & $2^{17}$ & $0.44918$                       & $0.46503$                  \\
            512  & $2^{18}$ & $0.45072$                       & $0.44533$                  \\
            1024 & $2^{19}$ & $0.45442$                       & $0.43561$                  \\ \bottomrule
        \end{tabular}                                                                                             \endgroup
    \end{center}
\end{table}

Table~\ref{tab:unstructured-weak} lists runtimes for increasing threads and
problem size on the GPU. Interpolate scales nearly perfectly with a difference
of $15\ms$ (\textasciitilde3\%) increase between the problem with 128 threads
and $2^{16}$ IB points and that with 1024 threads and $2^{19}$ points. Spread,
on the other hand, decreases in time as the problem size increases. This
speedup is artificial, and should not be expected in general. In the $n=2^{16}$
case, there is 1 IB point for every 4 grid cells, on average. When $n=2^{19}$,
the density increases to 2 for every grid cell. As a result, it becomes
increasingly unlikely to find a cell containing no IB points. This means that
writing the values to the output vector(s) becomes increasingly coalesced,
which, in turn, reduces the number of writes to global memory and vastly
improves the speed of the write overall. Typical use of the IB method does not
have IB points in every grid cell, but the recommendations that IB points on
connected structures be spaced $0.5h$--$h$ apart typically yields 1--4 IB
points in each occupied grid cell. We now consider a more typical use of the IB
method.


\subsection{Elastic objects}

We are motivated by the desire to simulate the motion of cells immersed in a fluid. Cells
are not randomly generated points, but cohesive structures, kept together by elastic
forces and the near-constant volume enclosed by their membranes. In this section, we
replace the randomly-placed IB points with points sampled on the surface of either a
sphere or an RBC. We track $n_d$ data sites per object, and interpolate fluid velocities
only to these points. We construct an RBF interpolant based on the positions of the data
sites and evaluate forces at $n_s$ sample sites, chosen so that neighboring sample sites
are initially within approximately $0.5h$ of each other. We spread forces from the sample
sites. It is generally the case that $n_s > n_d$, so that we interpolate to fewer points
than we spread from. In the parlance of Section \ref{sec:parallel}, $n_\gamma=n_d$ in the
context of interpolation, and $n_\gamma=n_s$ in the context of spreading. {\revb This violates
the transpose property of Equations~\eqref{eq:matrix-interp}--\eqref{eq:matrix-spread}.
Neither $\spread$ nor $\interp$ are constructed explicitly, so the algorithms presented
above work without modification. However, the $\interface_h$ argument will differ between
the spreading and interpolation algorithms.} Data and sample sites are generated using
the method described by~\cite{Shankar:2018ur}. In this case, we invoke the fluid solver,
so that as the object deforms, the force it imparts on the fluid will affect the fluid
velocity. The sphere and RBC are elastic, obeying the Skalak constitutive law
(\cite{Skalak:1973tp}) with shear modulus $2.5\times10^{-3}\dynpercm$ and bulk modulus
$2.5\times10^{-1}\dynpercm$. The RBC has the reference configuration given in~%
\cite{Omori:2012hw}:
\begin{equation}
    \begin{aligned}
        x(\theta,\,\varphi) &= R_0\cos\theta\cos\varphi, \\
        y(\theta,\,\varphi) &= R_0\sin\theta\cos\varphi, \\
        z(\theta,\,\varphi) &= R_0p(\cos \varphi) \sin\varphi,
    \end{aligned}
\end{equation}
where $\theta\in[-\pi,\,\pi)$, $\varphi\in[-\pi/2,\,\pi/2]$, $R_0=3.91\um$, and
$p(r)=0.105+r^2-0.56r^4$. These tests require a timestep of $k=0.1\us$ for stability.
With $\shear=1000\si{\per\second}$, an IB point requires at least 32 timesteps to transit
a grid cell, so unlike the tests using randomly placed IB points above, there will be
considerably more redundant computation. We first validate the fluid solver with the
elastic sphere before performing scaling tests, similar to those above, with RBCs.

\subsubsection{Convergence study}
To test the convergence of the fluid solver and cell representation, consider 
an object that obeys Skalak's law with the coefficients given above and is
spherical at rest. Deform the object from its rest configuration by stretching
it by a factor of $1.1$ in the $z$ direction and compressing it by a factor of
$1.1$ in the $y$ direction to maintain a fixed volume. For this test, $\shear$
is zero, so the fluid velocity is intially zero, and the object tends toward
its rest configuration over the course of simulation. We allow the object to
relax for $16\us$, and compare errors generated by successive grid refinements
of 16, 32, 64, and 128 points per $16\um$. For each grid, we use a fixed set of
$n_d=625$ data sites, sampled approximately uniformly on the surface of the
sphere, and choose $n_s$ so that sample sites are approximately uniform and
roughly $h/1.1$ apart, so that sample sites are roughly $h$ apart initially.
For $h=1\um$, $n_s=220$, and a refinement by a factor of 2 increases $n_s$ by a
factor of 4, for a maximum number of 14080 sample sites for these tests. A thin
interface which generates a force will cause a jump in the normal stress across
the interface, which the IB method may not recover. We therefore anticipate
first order convergence for the fluid velocity and data site positions.

\begin{table*}[ht]
    \caption{%
Convergence of the fluid velocity for a deformed sphere returning to its rest
configuration in a $16\um\times16\um\times16\um$ triply periodic domain at
$t=160\us$. The finest grid, with $h=0.125\um$ uses timestep $k=0.025\us$.
    }\label{tab:u-convergence}
    \begin{center}
        \begingroup
        \setlength{\tabcolsep}{9pt}
        \renewcommand{\arraystretch}{1.5}
        \begin{tabular}{cc|cc|cc}
                                                                                                                       \toprule
            $h$ ($\um$) & $k$ ($\us$) & $\|\u_h-\u_{0.5h}\|_2$ & order   & $\|\u_h-\u_{0.5h}\|_{\infty}$ & order    \\ \midrule
            $1.00$      & $1.6$       & $2.2274\times10^{-2}$  &         & $7.4187\times10^{-2}$         &          \\
            $0.50$      & $0.4$       & $4.3280\times10^{-4}$  & 5.71513 & $1.9083\times10^{-3}$         & 5.28079  \\
            $0.25$      & $0.1$       & $1.4684\times10^{-4}$  & 1.55951 & $1.0847\times10^{-3}$         & 0.81497  \\ \bottomrule
        \end{tabular}
        \endgroup
    \end{center}
\end{table*}

\begin{table*}[ht]
    \caption{%
Convergence of data sites for a deformed sphere returning to its rest
configuration in a $16\um\times16\um\times16\um$ triply periodic domain at
$t=16\us$. For each grid, we track 625 data sites on the sphere. The finest
grid, with $h = 0.125\um$ used $n_s=14080$ sample sites.
    }\label{tab:x-convergence}
    \begin{center}
        \begingroup
        \setlength{\tabcolsep}{9pt}
        \renewcommand{\arraystretch}{1.5}
        \begin{tabular}{cc|cc|cc}
                                                                                                                \toprule
            $h$ ($\um$) & $n_s$ & $\|\X_h-\X_{0.5h}\|_2$ & order   & $\|\X_h-\X_{0.5h}\|_{\infty}$ & order   \\ \midrule
            $1.00$      & 220   & $2.9611\times10^{-3}$  &         & $4.7812\times10^{-3}$         &         \\
            $0.50$      & 880   & $7.2997\times10^{-4}$  & 2.02020 & $1.1687\times10^{-3}$         & 2.03253 \\
            $0.25$      & 3520  & $3.5909\times10^{-4}$  & 1.02351 & $6.0956\times10^{-4}$         & 0.93902 \\ \bottomrule
        \end{tabular}
        \endgroup
    \end{center}
\end{table*}

Tables~\ref{tab:u-convergence} and~\ref{tab:x-convergence} show the convergence
of fluid velocity and data sites, respectively. To compute errors in the fluid
velocity, we construct a cubic spline from the velocities on the finer grid
and evaluate the spline at the grid points of the coarser grid. Each of the
interfaces is constructed with the same number of data sites, so the
coordinates from one simulation to another can be compared directly. We recover
approximately first order convergence for both fluid velocity and data sites
positions. Having established asymptotic convergence of our IB solver, we
continue by performing scaling tests with RBCs.

\subsubsection{Strong scaling}

We again wish to see how these algorithms can help speed up the runtime of
a fixed problem. Here, we consider tests with a single RBC and with 4 RBCs.
To construct the RBCs, we now use $n_d=864$ data sites, for an initial data
site spacing of approximately $1.6h$, and $n_s=8832$ sample sites per cell, for
an initial sample site spacing of approximately $0.5h$. We use a timestep of
$k=0.1\us$ to simulate the motion of the cells for $1\ms$.

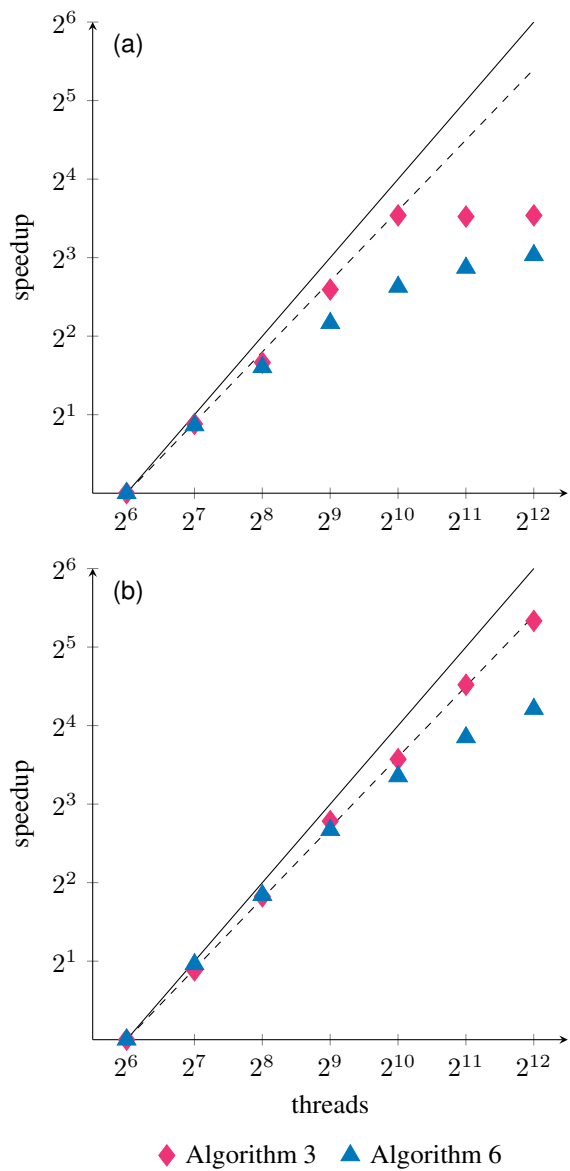
\begin{figure}[htbp]
\centering
\begin{tikzpicture}
\begin{groupplot}[
    group style={group name=rbc-strong, group size=1 by 2},
    height=0.45\textwidth,
    width=0.45\textwidth
]
\nextgroupplot[
        xmode=log,
        xmin=45.2548339959,
        xmax=5792.61875148,
        log basis x=2,
        ymode=log,
        ymax=64,
        log basis y=2,
        log origin=infty,
        width=0.45\textwidth,
        height=0.45\textwidth,
        axis lines=center,
        ylabel={speedup},
        xlabel near ticks,
        ylabel near ticks,
        legend style={at={(0.8, 0.2)}, anchor=center},
        legend cell align={left}
]
    \addplot+[only marks, mark=diamond*, color=tol/vibrant/magenta, mark options={scale=2, fill=tol/vibrant/magenta}] coordinates {%
        (64  , {0.01080/0.01080})
        (128 , {0.01080/0.00586})
        (256 , {0.01080/0.00341})
        (512 , {0.01080/0.00179})
        (1024, {0.01080/0.00093})
        (2048, {0.01080/0.00094})
        (4096, {0.01080/0.00093})
    }; \label{plot:rbc-interp}
    \addplot+[only marks, mark=triangle*, color=tol/vibrant/blue, mark options={scale=2, fill=tol/vibrant/blue}] coordinates {%
        (64  , {0.11888/0.11888})
        (128 , {0.11888/0.06535})
        (256 , {0.11888/0.03912})
        (512 , {0.11888/0.02649})
        (1024, {0.11888/0.01924})
        (2048, {0.11888/0.01627})
        (4096, {0.11888/0.01455})
    }; \label{plot:rbc-spread}
    \addplot+[no marks, dashed, black, domain=64:4096] {(x/64)^0.9};
    \addplot+[no marks, black, domain=64:4096] {(x/64)};
    \node [fill=white] at (rel axis cs: 0.075, 0.95) {\sffamily(a)};
\nextgroupplot[
        xmode=log,
        xmin=45.2548339959,
        xmax=5792.61875148,
        log basis x=2,
        ymode=log,
        ymax=64,
        log basis y=2,
        log origin=infty,
        width=0.45\textwidth,
        height=0.45\textwidth,
        axis lines=center,
        xlabel={threads},
        ylabel={speedup},
        xlabel near ticks,
        ylabel near ticks,
        legend style={at={(0.8, 0.2)}, anchor=center},
        legend cell align={left}
    ]
    \addplot+[only marks, mark=diamond*, color=tol/vibrant/magenta, mark options={scale=2, fill=tol/vibrant/magenta}] coordinates {%
        (64  , {0.04150/0.04150})
        (128 , {0.04150/0.02251})
        (256 , {0.04150/0.01172})
        (512 , {0.04150/0.00603})
        (1024, {0.04150/0.00349})
        (2048, {0.04150/0.00181})
        (4096, {0.04150/0.00103})
    };
    \addplot+[only marks, mark=triangle*, color=tol/vibrant/blue, mark options={scale=2, fill=tol/vibrant/blue}] coordinates {%
        (64  , {0.40148/0.40148})
        (128 , {0.40148/0.20628})
        (256 , {0.40148/0.11208})
        (512 , {0.40148/0.06313})
        (1024, {0.40148/0.03931})
        (2048, {0.40148/0.02785})
        (4096, {0.40148/0.02168})
    };
    \addplot+[no marks, dashed, black, domain=64:4096] {(x/64)^0.9};
    \addplot+[no marks, black, domain=64:4096] {(x/64)};
    \node [fill=white] at (rel axis cs: 0.075, 0.95) {\sffamily(b)};
\end{groupplot}
\path (rbc-strong c1r2.south west|-current bounding box.south)--
coordinate(legendpos) (rbc-strong c1r2.south east|-current bounding box.south);
\matrix[
    matrix of nodes,
    anchor=north,
    inner sep=0.2em,
  ]at([yshift=-1ex]legendpos)
  {
      \ref{plot:rbc-interp}& Algorithm \ref{algo:par-interp}&[5pt]
      \ref{plot:rbc-spread}& Algorithm \ref{algo:otf-spread} \\};
\end{tikzpicture}
\caption{%
    Speedup of Algorithms \ref{algo:par-interp} and \ref{algo:otf-spread} with
    increasing numbers of threads compared to 64 threads on the GPU for (a) 1
    and (b) 4 RBCs. Speedup is measured relative to the time taken for each
    algorithm using 64 threads on the GPU. Dashed lines indicate trends, and
    solid lines indicate ideal scaling.
}
\label{fig:str-strong}
\end{figure}

Figure \ref{fig:str-strong} shows the speedup observed with increasing threads
for 1 and 4 RBCs for 64--4096 threads on the GPU. We again see that the initial
speedup for interpolation is nearly linear with increased threads. In subfigure
\ref{fig:str-strong}(a), there is a sharp plateau that corresponds to every
data site having its own thread. In other words, there are more threads than
there is work to be done, since we track only 864 data sites for a single RBC.
Subfigure \ref{fig:str-strong}(b), on the other hand, has 3456 data sites, so
the trend continues for 512--4096 threads. In this case, we expect this graph
to plateau beyond 4096, when each data site has its own thread. However, we do
not expect for the trend to continue with more cells, as the presumed maximum
number of threads for interpolation is 4480, as discussed in Section
\ref{sec:unst-strong}. Comparing subfigure \ref{fig:str-strong}(a)
to (b), we see that the speedup in spreading is also dependent on the amount of
work. This indicates that, as with interpolation, the maximum speedup for
spreading is limited by hardware, rather than being a limitation of the
algorithm.

The trendlines for these tests indicate that increasing computing resources
by a factor of two decreases runtime by a factor of about 1.87 for these
algorithms. Again, this is merely an approximation as the sort and reduction
steps of the spreading algorithm are provided by {\thrust}, and therefore are
not limited to the listed number of threads. The similarity between the result
of the tests with RBCs and with randomly placed points indicates that the
distribution of points does not have a marked impact on the efficacy of the
parallelization for a fixed problem. We now see if the same holds for weak
scaling tests.


\subsubsection{Weak scaling}

To see how the algorithms scale given more computing resources, we increase the
number of cells in the domain and the number of threads proportionally. We
construct each cell with $n_d=864$ data sites and $n_s=8832$ sample sites, as
before. We place between 1 and 8 cells in the domain, while threads increase
from 64 to 512. Using a timestep of $k = 0.1\us$, we simulate the motion of
these cells for $1\ms$.

In Section~\ref{sec:unst-weak}, we observe that, as a side-effect of increasing
point density per grid cell, runtime for spreading decreases as the number of
points and threads increases. Here, the cells are initially far enough part as
to not have any overlapping support points.  As a result, while individual grid
cells may contain several IB points, average point density is still low, so we
do not expect to see the same reduction in runtime as observed previously.

\begin{table}[t]
    \caption{%
Weak scaling results for interpolate and spreading for increasing numbers of
RBCs (cells column) and threads. Each RBC has $n_d = 864$ and $n_s = 8832$.
Average time per call is reported in seconds. $N$ is the number of samples
taken.}\label{tab:str-weak}
    \begin{center}
        \begingroup
        \setlength{\tabcolsep}{9pt}
        \renewcommand{\arraystretch}{1.5}
        \begin{tabular}{ccccc}
                                                                                          \toprule
            $p$ & cells & \titletable{interpolate}{20000} & \titletable{spread}{10000} \\ \midrule
            64  & 1     & $0.01079$                       & $0.11881$                  \\
            128 & 2     & $0.01165$                       & $0.11219$                  \\
            256 & 4     & $0.01171$                       & $0.11214$                  \\
            512 & 8     & $0.01199$                       & $0.11354$                  \\ \bottomrule
        \end{tabular}
        \endgroup
    \end{center}
\end{table}

Table~\ref{tab:str-weak} shows the runtimes for increasing number of RBCs and
threads. We observe the near-perfect scaling we saw with random IB points. For
both interpolate and spread, we see that the runtimes are nearly constant: the
slowest and fastest times differ by less than $2\ms$ and $7\ms$, for
interpolation and spread, respectively. After an initial drop in runtime
between one RBC with 64 threads and two RBCs and 128 threads, runtimes even off
and begin to increase, in contrast with the results in Section~%
\ref{sec:unst-weak}.


\section{Summary} \label{sec:summary}

We presented {\revb Algorithm~\ref{algo:par-interp} for interpolation in parallel and introduced
a novel parallel algorithm, Algorithm~\ref{algo:par-spread},} for spreading forces from
time-dependent interfaces in the immersed boundary method. These algorithms have a
runtime that is independent of the Eulerian grid. This makes the parallelization useful
for cases where IB points inhabit only a small percentage of the Eulerian grid cells. We
also introduced two variants for spread which trade off dependence on the Eulerian grid,
in the form of a few vector adds and memory allocation, for improved runtimes for small
enough grids.

These algorithms exhibit nearly ideal scaling, on both the CPU and GPU, for problems of
fixed size and increasing number of threads, as well as for problems of increasing size
and number of threads.  We observed that on the GPU, larger problem sizes led to higher
speedup plateaus, indicating that larger problems on more capable hardware will achieve
even larger speedups than presented here.

{\revaa It is instructive to compare Algorithm~\ref{algo:par-spread}, which computes one value of
$\Dirac_h$ for each Lagrangian point per thread synchronization, with the parallelization
of~\cite{McQueen:1997kw}, which computes all values for $\Dirac_h$ for a subset of the
Lagrangian points each thread synchronization. Computing all values of $\Dirac_h$ at the
same time allows for minimizing calls to $\kernel$, while processing all Lagrangian
points simultaneously eliminates dependence on the background grid. It is straightforward
to construct a pathological case for which McQueen and Peskin's scheme performs poorly:
if all of the Lagrangian points lie within the same $s\times s$ columns of Eulerian grid
cells, then a single thread processes all of the Lagrangian points, reducing the scheme
to a serial one. The fine-grained approach in Algorithm~\ref{algo:par-spread} lacks this
pathology, making it more robust to IB point configurations.

Though efficient load-balancing for Lagrangian work requires special techniques
(\cite{Givelberg:2006gg,Bhalla:2019vb}), the worst-case scenario for distributed memory
parallelizations has every Lagrangian object mapped to the same device
(\cite{Erickson:2010uzba}). For that reason, single-node results are still relevant.
Nevertheless, we anticipate that our algorithms will perform well in multi-node systems
as they do not rely on partitioning the Eulerian or Lagrangian points.} Because
information about the Eulerian grid is encoded by the functions $\key$ and $\#$, an
optimized distributed-memory algorithm may need to use a different $\key$ or $\#$
function on each node to account for Eulerian grid partitioning. As simulations become
more computationally intensive, such algorithms will become more necessary. To that end,
future work involves adapting our algorithms for multi-node and multi-GPU systems.


\begin{acks}
The authors acknowledge useful discussions with Dr. Aaron Barrett.
\end{acks}

\begin{dci}
The authors have no conflicts of interest to declare.
\end{dci}

\begin{funding}
The author(s) disclosed receipt of the following financial support for the research,
authorship, and/or publication of this article: This work was supported by the National
Science Foundation [grant numbers DMS-1521748, DMS-176898, CISE CCF-1714844].
\end{funding}

\bibliography{ib-paper}  

\begin{thebibliography}{28}
\providecommand{\natexlab}[1]{#1}
\providecommand{\url}[1]{\texttt{#1}}
\providecommand{\urlprefix}{URL }
\expandafter\ifx\csname urlstyle\endcsname\relax
  \providecommand{\doi}[1]{DOI:\discretionary{}{}{}#1}\else
  \providecommand{\doi}{DOI:\discretionary{}{}{}\begingroup
  \urlstyle{rm}\Url}\fi

\bibitem[{Bhalla et~al.(2019)Bhalla, Nangia, Dafnakis, Bracco and
  Mattiazzo}]{Bhalla:2019vb}
Bhalla APS, Nangia N, Dafnakis P, Bracco G and Mattiazzo G (2019) {Simulating
  water-entry/exit problems using Eulerian-Lagrangian and fully-Eulerian
  fictitious domain methods within the open-source IBAMR library}.
\newblock \emph{arXiv.org} : 1--32.

\bibitem[{Brown et~al.(2001)Brown, Cortez and Minion}]{Brown:2001bq}
Brown DL, Cortez R and Minion ML (2001) {Accurate Projection Methods for the
  Incompressible Navier{\textendash}Stokes Equations}.
\newblock \emph{Journal of Computational Physics} 168(2): 464--499.

\bibitem[{Chuang et~al.(2018)Chuang, Mesnard, Krishnan and
  Barba}]{Chuang:2018ej}
Chuang PY, Mesnard O, Krishnan A and Barba LA (2018) {PetIBM: toolbox and
  applications of the immersed-boundary method on distributed-memory
  architectures}.
\newblock \emph{Journal of Open Source Software} 3(25): 558.

\bibitem[{Erickson(2010)}]{Erickson:2010uzba}
Erickson LC (2010) \emph{{Blood Flow Dynamics: a Lattice Boltzmann Immersed
  Boundary Approach}}.
\newblock PhD Thesis, University of Utah, Salt Lake City.

\bibitem[{Fai et~al.(2013)Fai, Griffith, Mori and Peskin}]{Fai:2013do}
Fai TG, Griffith BE, Mori Y and Peskin CS (2013) {Immersed Boundary Method for
  Variable Viscosity and Variable Density Problems Using Fast
  Constant-Coefficient Linear Solvers I: Numerical Method and Results}.
\newblock \emph{SIAM Journal on Scientific Computing} 35(5): B1132--B1161.

\bibitem[{Fuselier et~al.(2013)Fuselier, Hangelbroek, Narcowich, Ward and
  Wright}]{Fuselier:2013coba}
Fuselier EJ, Hangelbroek T, Narcowich FJ, Ward JD and Wright GB (2013) {Kernel
  based quadrature on spheres and other homogeneous spaces}.
\newblock \emph{Numerische Mathematik} 127(1): 57--92.

\bibitem[{Givelberg and Yelick(2006)}]{Givelberg:2006gg}
Givelberg E and Yelick K (2006) {Distributed Immersed Boundary Simulation in
  Titanium}.
\newblock \emph{SIAM Journal on Scientific Computing} 28(4): 1361--1378.

\bibitem[{Griffith(2009)}]{Griffith:2009gg}
Griffith BE (2009) {An accurate and efficient method for the incompressible
  Navier{\textendash}Stokes equations using the projection method as a
  preconditioner}.
\newblock \emph{Journal of Computational Physics} 228(20): 7565--7595.

\bibitem[{Griffith(2011)}]{Griffith:2011gi}
Griffith BE (2011) {Immersed boundary model of aortic heart valve dynamics with
  physiological driving and loading conditions}.
\newblock \emph{International Journal for Numerical Methods in Biomedical
  Engineering} 28(3): 317--345.

\bibitem[{Griffith et~al.(2007)Griffith, Hornung, McQueen and
  Peskin}]{Griffith:2007do}
Griffith BE, Hornung RD, McQueen DM and Peskin CS (2007) {An adaptive, formally
  second order accurate version of the immersed boundary method}.
\newblock \emph{Journal of Computational Physics} 223(1): 10--49.

\bibitem[{Griffith et~al.(2010)Griffith, Hornung, McQueen and
  Peskin}]{Griffith:2007uk}
Griffith BE, Hornung RD, McQueen DM and Peskin CS (2010) {Parallel and Adaptive
  Simulation of Cardiac Fluid Dynamics}.
\newblock In: Parashar M and Li X (eds.) \emph{Advanced Computational
  Infrastructures for Parallel and Distributed Adaptive Applications}. Hoboken,
  NJ, USA: John Wiley {\&} Sons, Inc., pp. 105--130.

\bibitem[{Griffith and Luo(2017)}]{Griffith:2017id}
Griffith BE and Luo X (2017) {Hybrid finite difference/finite element immersed
  boundary method}.
\newblock \emph{International Journal for Numerical Methods in Biomedical
  Engineering} 33: e2888.

\bibitem[{Griffith and Patankar(2020)}]{Griffith:2020hi}
Griffith BE and Patankar NA (2020) {Immersed Methods for
  Fluid{\textendash}Structure Interaction}.
\newblock \emph{Annual Review of Fluid Mechanics} 52(1): 421--448.

\bibitem[{Harlow and Welch(1965)}]{Welch:1965jv}
Harlow FH and Welch JE (1965) {Numerical Calculation of Time-Dependent Viscous
  Incompressible Flow of Fluid with Free Surface}.
\newblock \emph{Physics of Fluids} 8(12): 2182--2189.

\bibitem[{Kassen(2021)}]{ibcu}
Kassen A (2021) sonwell/ib.cu.
\newblock \urlprefix\url{https://github.com/sonwell/ib.cu}.

\bibitem[{Layton et~al.(2011)Layton, Krishnan and Barba}]{Layton:2011um}
Layton SK, Krishnan A and Barba LA (2011) {cuIBM {\textendash} A
  GPU-accelerated Immersed Boundary Method }.
\newblock \emph{arXiv.org} .

\bibitem[{Maxian et~al.(2018)Maxian, Kassen and Strychalski}]{Maxian:2018ek}
Maxian O, Kassen AT and Strychalski W (2018) {A continuous energy-based
  immersed boundary method for elastic shells}.
\newblock \emph{Journal of Computational Physics} 371: 333--362.

\bibitem[{McQueen and Peskin(1997)}]{McQueen:1997kw}
McQueen DM and Peskin CS (1997) {Shared-Memory Parallel Vector Implementation
  of the Immersed Boundary Method for the Computation of Blood Flow in the
  Beating Mammalian Heart}.
\newblock \emph{The Journal of Supercomputing} 11(3): 213--236.

\bibitem[{Mesnard and Barba(2017)}]{Mesnard:2017te}
Mesnard O and Barba LA (2017) {Reproducible and Replicable Computational Fluid
  Dynamics}.
\newblock \emph{Computing in Science Engineering} 19(4): 44--55.

\bibitem[{Mittal and Iaccarino(2005)}]{Iaccarino:2005ii}
Mittal R and Iaccarino G (2005) {Immersed Boundary Methods}.
\newblock \emph{Annual Review of Fluid Mechanics} 37(1): 239--261.

\bibitem[{Omori et~al.(2012)Omori, Ishikawa, Barth{\`e}s-Biesel, Salsac, Imai
  and Yamaguchi}]{Omori:2012hw}
Omori T, Ishikawa T, Barth{\`e}s-Biesel D, Salsac AV, Imai Y and Yamaguchi T
  (2012) {Tension of red blood cell membrane in simple shear flow}.
\newblock \emph{Physical Review E} 86(5): 056321--056329.

\bibitem[{Patel(2012)}]{Patel:2012tc}
Patel S (2012) \emph{{aeroCuda: The GPU-Optimized Immersed Solid Code}}.
\newblock PhD Thesis, Harvard, Harvard University.

\bibitem[{Peskin(1972)}]{Peskin:1972wa}
Peskin CS (1972) \emph{{Flow patterns around heart valves: a digital computer
  method for solving the equations of motion}}.
\newblock PhD Thesis, Yeshiva University, New York.

\bibitem[{Peskin(2002)}]{Peskin:2002go}
Peskin CS (2002) {The immersed boundary method}.
\newblock \emph{Acta Numerica} 11: 479--517.

\bibitem[{Shankar et~al.(2018)Shankar, Kirby and Fogelson}]{Shankar:2018ur}
Shankar V, Kirby RM and Fogelson AL (2018) {Robust Node Generation for Meshfree
  Discretizations on Irregular Domains and Surfaces}.
\newblock \emph{SIAM Journal on Scientific Computing} 40(4): A2584--A2608.

\bibitem[{Shankar et~al.(2015)Shankar, Wright, Kirby and
  Fogelson}]{Shankar:2015km}
Shankar V, Wright GB, Kirby RM and Fogelson AL (2015) {Augmenting the immersed
  boundary method with Radial Basis Functions (RBFs) for the modeling of
  platelets in hemodynamic flows}.
\newblock \emph{International Journal for Numerical Methods in Fluids} 79(10):
  536--557.

\bibitem[{Skalak et~al.(1973)Skalak, Tozeren, Zarda and Chien}]{Skalak:1973tp}
Skalak R, Tozeren A, Zarda RP and Chien S (1973) {Strain Energy Function of Red
  Blood Cell Membranes}.
\newblock \emph{Biophysical Journal} 13: 245--264.

\bibitem[{Wesseling(2001)}]{Wesseling:2001ci}
Wesseling P (2001) \emph{{Principles of Computational Fluid Dynamics}},
  \emph{Springer Series in Computational Mathematics}, volume~29.
\newblock Berlin, Heidelberg: Springer Berlin Heidelberg.

\end{thebibliography}
\bibliographystyle{SageH}

\end{document}